\documentclass[twocolumn]{aastex631}
\usepackage{amsmath,amstext,amssymb}
\usepackage{verbatim}
\usepackage{soul}
\usepackage[T1]{fontenc}
\usepackage[figure,figure*]{hypcap}
\usepackage{rotating}
\usepackage{multirow}
\usepackage{booktabs}
\usepackage{enumitem}

\newcommand{\approptoinn}[2]{\mathrel{\vcenter{
 \offinterlineskip\halign{\hfil$##$\cr
 #1\propto\cr\noalign{\kern2pt}#1\sim\cr\noalign{\kern-2pt}}}}}

\newcommand{\be}{\begin{equation}}
\newcommand{\ee}{\end{equation}}

\usepackage{xspace}
\newcommand\codename[1]{\textsc{#1}\xspace}

\def\athenapk/{\codename{AthenaPK}}
\def\flash/{\codename{FLASH}}

\begin{document}

\title{An Improved Fit to the Density Fluctuations in Supersonic Isothermal Turbulence}
\shorttitle{An Improved Fit to Supersonic Turbulence}
\shortauthors{Scannapieco, Br\"uggen, Grete, \& Pan}

\author[0000-0002-3193-1196]{Evan Scannapieco}
\affil{School of Earth \& Space Exploration, 
Arizona State University, 781 Terrace Mall, Tempe, AZ 85287, USA; evan.scannapieco@asu.edu}

\author[0000-0002-3369-7735]{Marcus Br\"uggen}
\affil{University of Hamburg, Hamburger Sternwarte, Gojenbergsweg 112, 21029, Hamburg, Germany}

\author[0000-0003-3555-9886]{Philipp Grete}
\affil{University of Hamburg, Hamburger Sternwarte, Gojenbergsweg 112, 21029, Hamburg, Germany}

\author[0000-0002-0502-8593]{Liubin Pan}
\affil{School of Physics and Astronomy, Sun Yat-sen University, 2 Daxue Road, Zhuhai, Guangdong, 519082, People's Republic of China; panlb5@mail.sysu.edu.cn}

\begin{abstract}

The density distribution of supersonic isothermal turbulence plays a critical role in many astrophysical systems. It is commonly approximated by a lognormal distribution with a variance of $\sigma_{s,{\rm V}}^2 \approx \ln(1 + b^2 M_{\rm V}^2),$ where $s \equiv \ln \rho/\rho_0,$ $M_{\rm V}$ is the rms volume-weighted Mach number, and $b$ is a parameter that depends on the driving mechanism, which can be solenoidal (divergence-free), compressive (curl-free), or a mix of both. However, this fit neglects the driving correlation time, $\tau_{\rm a}$, which plays a key role when compressive driving is significant. Here we conduct turbulence simulations spanning a wide range of Mach numbers, driving mechanisms, and $\tau_{\rm a}$ values. In the compressive case,  $\sigma_{s,{\rm V}}^2$ is not well fit by the standard expression.  Instead, it scales approximately linearly with $M_{\rm V},$ and its dependence on $\tau_{\rm a}$ is $\sigma_{s,{\rm V}}^2 \approx M_{\rm V} [1 + \frac{2}{3}(1 + \lambda_{\rm a})\Theta(1 + \lambda_{\rm a})]$, where $\lambda_{\rm a} \equiv \ln(\tau_{\rm a}/\tau_{\rm e})$, $\tau_{\rm e}$ is the eddy turnover time, and $\Theta$ is the Heaviside step function. Mixed-driven turbulence shows a weak dependence on $\tau_{\rm a},$  and for solenoidally-driven turbulence, $\sigma_{s,{\rm V}}^2 \approx \frac{1}{3}M_{\rm V}$, which is consistent with the standard expression when $M_{\rm V} \lesssim 8.$  The volume-weighted mean and skewness also show systematic trends with $M_{\rm V}$ and $\tau_{\rm a}$, deviating from lognormal expectations. The mass-weighted density distribution displays significant broadening and skewness in compressively-driven cases, especially at large $\tau_{\rm a}/\tau_{\rm e}$. These results provide a refined framework for modeling astrophysical turbulence.
 
\end{abstract}


\section{Introduction}

Turbulence plays a fundamental role in shaping numerous astrophysical systems, from molecular clouds and the interstellar medium to galaxy clusters and the circumgalactic medium \cite[e.g.][]{Brandenburg95, Kim03, Moesta15, Walch15, Zhuravleva18, Buie20, Ostriker22, Rosotti23}. In each of these systems, the dynamics, structure, and evolution depend not only on overall properties, such as magnetic field strength and effective equation of state, \cite[e.g.][]{Zweibel95, Krumholz06, Hennebelle09, Tasker09, Hopkins13b, Li15, Federrath15, Xu19}, but also on the features of the turbulence itself, such as the Mach number and driving mechanism \cite[e.g.][]{Padoan97, Ostriker2001, Federrath08, Burkhart09, Pan19}.

In nature, turbulent motions are driven by a variety of processes. These include shear and the magnetorotational instability, which drive mostly solenoidal (or divergence-free) motions \citep[e.g.,][]{Kim03, Tamburro09, Sur16, Federrath16}, as well as gravitational collapse, tidal forces, thermal instability, and feedback, which drive mostly compressive (or curl-free) motions \citep[e.g.,][]{VazquezSemadeni98, Dobbs08, Klessen10, Robertson12, McKee89, Goldbaum11, Peters11, Renaud15}. There have also been studies of cases in which the driving agents and the generated modes are less clearly correlated, such as the amplitude of solenoidal modes increasing in collapsing distributions \citep{2021ApJ...915..107H,2023A&A...672A.193F} and supernovae giving rise to solenoidal modes when propagating through inhomogeneous media \citep{2004ApJ...617..339B,Padoan16,Kobayashi22}.

In simulations, driving processes are often approximated by random accelerations, modeled either as a static pattern, a pattern that changes every timestep, or, more commonly, as an Ornstein-Uhlenbeck (OU) process with a finite autocorrelation timescale, $\tau_{\rm a}$ \citep[e.g.][]{Eswaran88, Schmidt09}. The driving strength then sets the Mach number, whereas the driving pattern determines the solenoidal to compressive ratio of large-scale motions, which cascade toward small scales, forming shocks and density structures through nonlinear processes.

These structures also generally depend on the equation of state. The 1D simulations of \citet{Passot98}, for example, showed that the volume-weighted probability density function (PDF) of the density fluctuations approaches a power-law at high densities for polytropic indexes smaller than unity, and a power-law at low densities for polytropic indexes greater than unity. In constrast, \citet{Federrath15} found that in 3D, the PDF did not approach a power-law at high densities for polytropic indexes below 1.

In cases in which the effective equation of state is approximately isothermal, the volume-weighted PDF of the logarithmic density is often approximated by a Gaussian,
\be
P_{\rm V}(s) ds \approx \frac{1}{\sqrt{2 \pi \sigma_{s,{\rm V}}^2}} {\rm exp} \left[ - \frac{(s - \left<s \right>_{\rm V})^2} {2 \sigma_{s,{\rm V}}^{2}} \right] ds,
\label{eq:PDF}
\ee
where $s \equiv \ln(\rho/\rho_0)$, $\rho_0$ is the mean density, and the mean value of $s$ is related to the variance as $\left<s \right>_{\rm V}= - \sigma_{s,{\rm V}}^2/2$ by mass conservation \citep{VazquezSemadeni94,Padoan97,Federrath10,Padoan11}.

Several studies have sought to capture this distribution and its dependence on the Mach number and the driving mechanism through a fit to $\sigma_{s, V}^2$. The most widely applied such fit is
\be
\sigma_{s, V}^2 \approx \ln(1 + b_s^2 M_{\rm V}^2),
\label{eq:standardfit}
\ee
where $M_{\rm V}$ is the rms volume-weighted Mach number, and $b_s$ is a parameter that depends on the nature of the turbulent forcing, which is taken to be $b_s\approx$ $\frac{1}{3}$ in the solenoidal case and $b_s\approx 1$ in the compressive case \citep{Padoan97, Ostriker2001, MacLow05, Glover07, Lemaster08, Price11}.  Note that here we use the subscript $s$ to denote that this is a fit to the distribution of log density rather than to the density itself. This functional form is built on the assumption of a lognormal distribution and a relation between Mach number and density variance of the form $\sigma_{\rho,{\rm V}}^2 \propto M_{\rm V}^2,$ which is motivated by the fact that the density contrast behind an isothermal shock is proportional to $M_{\rm V}^2,$ but the dense shocked gas occupies only a fraction $M_{\rm V}^2$ of the original volume \citep{Padoan97}. A more physical explanation of the $\sigma_{\rho,{\rm V}}^2 \propto M_{\rm V}^2$ scaling, based on an exact result derived from the hydrodynamical equations, was put forward in \citet{Pan22}.

However, there are several underlying problems with adopting eqs.\ (\ref{eq:PDF}) and (\ref{eq:standardfit}). Although $\sigma_{\rho,{\rm V}}^2$ is expected to be proportional to $M_{\rm V}^2$ in the presence of a single isothermal shock, the density distribution in a turbulent distribution is instead set by a balance between stochastic compressions/expansions, which broaden the PDF, and the acceleration/deceleration of shocks by density gradients, which tends to narrow the PDF \citep{Paper1}. Secondly, while eq.\ (\ref{eq:PDF}) provides an approximate shape of $P_{\rm V},$ more detailed measurements have shown that the distribution is significantly skewed toward low densities, particularly at high Mach numbers and in compressively-driven turbulence \citep{Federrath08, Schmidt09, Konstandin12, Hopkins13a, Federrath13, Squire17, Mocz19}, meaning that $\sigma_{s, V}^2 \approx \ln(1 + b_s^2 M_{\rm V}^2)$ may not follow from $\sigma_{\rho,{\rm V}}^2 \propto M_{\rm V}^2.$

This has been studied by \cite{Squire17}, who presented an improved fit in the form of a compound log-Poisson model for the PDF that treats density variations as discrete multiplicative jumps, with sizes drawn from an exponential distribution. In further work, \cite{Rabatin23a} presented a shock model for the density PDF, modeling each gas parcel as experiencing a finite number of shocks before relaxing to the mean density. Compared with $M_{\rm V}=0.1-25$  simulations with various forcings (solenoidal, compressive, mixed),  their model fits PDFs up to an order of magnitude better than the standard lognormal model \citep[see also][]{Rabatin23b}.

However, the Mach number and the mix of compressive and solenoidal modes are not the only two parameters that determine the density distribution of isothermal turbulence. In \cite{Paper2}, we showed that when compressive driving is significant, the correlation time of driving accelerations, $\tau_{\rm a},$ also plays a critical role. When $\tau_{\rm a}$ is comparable to or larger than the eddy turnover time, $\tau_{\rm e}$, compressive driving produces large, low-density voids, leading to a broader and more skewed density PDF. In contrast, for $\tau_{\rm a} \ll \tau_{\rm e}$, these voids are suppressed, resulting in a narrower, more symmetric PDF. 

A possible link between $\tau_{\rm a}$ and the low-density tail of the PDF was mentioned by \citet{Konstandin12} as a source of non-Gaussian wings at high Mach numbers,  but this was not quantitatively analyzed. \citet{Alvelius1999} showed analytically that when $\tau_{\rm a}/\tau_{\rm e}$ is large, the acceleration field leaves a lasting imprint on the flow because it evolves more slowly than the dynamical time of the flow. Similarly, \citet{Grete2018} showed that  $\delta$-in-time forcing also impacts the compressible turbulence dynamics in an unphysical way. Together, these findings illustrate that the density structure cannot be reliably captured by without accounting for the driving correlation time. 

Here, we address this issue by conducting a suite of simulations to measure how the PDF of $s$ and its associated statistics vary over a 
parameter space that includes solenoidal, mixed, and compressive driving, a wide range of Mach numbers, and driving times ranging from from $\tau_{\rm a} \ll \tau_{\rm e}$ to $\tau_{\rm a} \gg \tau_{\rm e}.$ We then use these measurements to provide empirical relations for the mean and variance of the volume-weighted PDFs as functions of the flow properties.

Our results serve two complementary purposes. First, they test the physical assumption underlying eq.\ (\ref{eq:standardfit}), that $P(s)$ is approximately Gaussian and that the relation between Mach number and density variance is approximately $\sigma_{\rho,{\rm V}}^2 \propto M_{\rm V}^2.$ Second, they provide a more accurate characterization of the density distribution for future observational and theoretical studies \citep[e.g.][]{Banda-Barragan21,Sharda22}. By accounting for the previously-neglected dependence on $\tau_{\rm a}$, these formulae enable more realistic modeling of systems in which compressive driving plays a significant role.

The structure of this paper is as follows: In \S2, we describe the numerical methods and parameters spanned by the simulation suite. In \S3, we present our results, including the measured PDFs and their dependence on the driving mechanism, Mach number, and correlation time of the driving accelerations. We summarize our conclusions in \S4.

\section{Simulations}

\subsection{Methods}
\label{sec:Sims}

 To generate a suite of simulations of supersonic, isothermal turbulence, we adopted the methodology outlined in \cite{Paper1} and \cite{Paper2}. All simulations were conducted in a periodic domain of size \( L_{\rm box} \), within which we numerically solved the hydrodynamic equations under the influence of a stochastic driving force. The governing equations for mass and momentum conservation in this case are
 \begin{equation}
\frac{\partial \rho}{\partial t} + \frac{\partial \rho v_i}{\partial x_i} = 0,
\label{eq:continuity}
\end{equation} 
and 
\begin{equation}
  \frac{\partial \rho v_i}{\partial t} + \frac{\partial (\rho v_i v_j + \delta_{ij}p - \sigma_{ij})}{\partial x_j} = 
 \rho a_{i} (\mathbf{x}, t),
\label{eq:velocity}
\end{equation} 
where $p(\mathbf{x}, t)$ is the pressure, $\sigma_{ij}$ is the viscous stress tensor, and $\mathbf{a}(\mathbf{x}, t)$ is the driving force. For an ideal gas, the shear viscosity is $\sigma_{ij} = \rho \nu ( \partial_i v_j + \partial_j v_i - \frac{2}{3} \partial_k v_k \delta_{ij}) $ where $\nu$ is the kinematic viscosity. 

\begin{table*}[t]\centering\begin{tabular}{lccccccccccccc}\toprule
& \multicolumn{3}{c}{Input parameters} & \multicolumn{10}{c}{Simulation properties} \\
\cmidrule(lr{.75em}){2-4} \cmidrule(lr{.75em}){5-14}
\texttt{Id} & $\zeta$ & $a$ & $\tau_\mathrm{a}$ & $M_\mathrm{V}$ & $M_\mathrm{M}$ & $\tau_\mathrm{e}$ & ln$(\tau_\mathrm{a} / \tau_\mathrm{e})$ & $r_\mathrm{cs}$ & $\nu_{\rm eff}~[10^{-4}]$ & $\eta/\Delta_x$ & $\lambda/\Delta_x$ & $\mathrm{Re}_{\lambda}$ & $\mathrm{Re}_{\rm int}$ \\ \midrule
$\mathtt{Ms1.9\_C\_\lambda-3.1}$ & 0.0 & 35 & 0.01 & 1.9 & 1.8 & 0.21 & -3.07 & 0.81 & 8.4 & 1.37 & 52 & 230 & 900\\
$\mathtt{Ms1.9\_C\_\lambda-1.9}$ & 0.0 & 25 & 0.03 & 1.9 & 1.8 & 0.21 & -1.94 & 0.81 & 8.2 & 1.34 & 51 & 230 & 900\\
$\mathtt{Ms2.0\_C\_\lambda-0.7}$ & 0.0 & 25 & 0.10 & 2.0 & 1.8 & 0.19 & -0.67 & 0.80 & 7.8 & 1.33 & 47 & 240 & 1000\\
$\mathtt{Ms1.8\_C\_\lambda+0.3}$ & 0.0 & 25 & 0.30 & 1.8 & 1.4 & 0.21 & 0.34 & 0.80 & 7.4 & 1.49 & 46 & 220 & 900\\
$\mathtt{Ms1.4\_C\_\lambda+1.4}$ & 0.0 & 25 & 1.00 & 1.4 & 1.0 & 0.26 & 1.36 & 0.80 & 7.7 & 1.92 & 46 & 170 & 700\\
$\mathtt{Ms1.2\_C\_\lambda+2.3}$ & 0.0 & 25 & 3.00 & 1.2 & 0.7 & 0.31 & 2.28 & 0.82 & 7.7 & 2.34 & 49 & 150 & 600\\[0.3em]
$\mathtt{Ms2.8\_C\_\lambda-2.7}$ & 0.0 & 60 & 0.01 & 2.8 & 2.6 & 0.15 & -2.69 & 0.78 & 8.7 & 1.08 & 46 & 290 & 1300\\
$\mathtt{Ms3.0\_C\_\lambda-1.5}$ & 0.0 & 50 & 0.03 & 3.0 & 2.8 & 0.13 & -1.49 & 0.76 & 8.6 & 1.04 & 44 & 300 & 1400\\
$\mathtt{Ms3.0\_C\_\lambda-0.3}$ & 0.0 & 50 & 0.10 & 3.0 & 2.5 & 0.13 & -0.27 & 0.77 & 8.2 & 1.08 & 42 & 300 & 1500\\
$\mathtt{Ms2.7\_C\_\lambda+0.8}$ & 0.0 & 50 & 0.30 & 2.7 & 1.9 & 0.14 & 0.76 & 0.78 & 7.4 & 1.20 & 41 & 290 & 1400\\
$\mathtt{Ms2.4\_C\_\lambda+1.9}$ & 0.0 & 50 & 1.00 & 2.4 & 1.3 & 0.15 & 1.88 & 0.78 & 7.7 & 1.59 & 42 & 260 & 1200\\
$\mathtt{Ms2.1\_C\_\lambda+2.9}$ & 0.0 & 50 & 3.00 & 2.1 & 1.1 & 0.17 & 2.89 & 0.79 & 7.9 & 1.82 & 41 & 220 & 1000\\[0.3em]
$\mathtt{Ms4.7\_C\_\lambda-2.1}$ & 0.0 & 140 & 0.01 & 4.7 & 4.4 & 0.09 & -2.15 & 0.71 & 9.1 & 0.79 & 39 & 390 & 2100\\
$\mathtt{Ms4.4\_C\_\lambda-1.1}$ & 0.0 & 100 & 0.03 & 4.4 & 3.9 & 0.09 & -1.10 & 0.73 & 8.7 & 0.83 & 40 & 400 & 2100\\
$\mathtt{Ms4.3\_C\_\lambda+0.1}$ & 0.0 & 100 & 0.10 & 4.3 & 3.4 & 0.09 & 0.11 & 0.73 & 8.4 & 0.89 & 38 & 390 & 2000\\
$\mathtt{Ms4.0\_C\_\lambda+1.2}$ & 0.0 & 100 & 0.30 & 4.0 & 2.5 & 0.10 & 1.15 & 0.75 & 8.2 & 1.07 & 38 & 360 & 1800\\
$\mathtt{Ms4.0\_C\_\lambda+2.4}$ & 0.0 & 100 & 1.00 & 4.0 & 1.7 & 0.09 & 2.38 & 0.76 & 7.2 & 1.34 & 40 & 430 & 2100\\
$\mathtt{Ms3.6\_C\_\lambda+3.4}$ & 0.0 & 100 & 3.00 & 3.6 & 1.2 & 0.10 & 3.39 & 0.77 & 7.6 & 1.59 & 39 & 360 & 1700\\[0.3em]
$\mathtt{Ms7.1\_C\_\lambda-1.7}$ & 0.0 & 280 & 0.01 & 7.1 & 6.9 & 0.06 & -1.72 & 0.64 & 10.1 & 0.63 & 34 & 470 & 2800\\
$\mathtt{Ms6.6\_C\_\lambda-0.7}$ & 0.0 & 200 & 0.03 & 6.6 & 5.7 & 0.06 & -0.70 & 0.68 & 9.7 & 0.69 & 35 & 470 & 2700\\
$\mathtt{Ms6.3\_C\_\lambda+0.5}$ & 0.0 & 200 & 0.10 & 6.3 & 4.8 & 0.06 & 0.48 & 0.70 & 8.7 & 0.72 & 35 & 510 & 2900\\
$\mathtt{Ms6.1\_C\_\lambda+1.6}$ & 0.0 & 200 & 0.30 & 6.1 & 3.7 & 0.06 & 1.57 & 0.71 & 8.5 & 0.85 & 35 & 490 & 2700\\
$\mathtt{Ms6.2\_C\_\lambda+2.8}$ & 0.0 & 200 & 1.00 & 6.2 & 2.8 & 0.06 & 2.79 & 0.72 & 8.0 & 1.04 & 37 & 560 & 3000\\
$\mathtt{Ms6.5\_C\_\lambda+3.9}$ & 0.0 & 200 & 3.00 & 6.5 & 1.8 & 0.06 & 3.94 & 0.71 & 7.7 & 1.32 & 39 & 640 & 3200\\[0.3em]
$\mathtt{Ms10.4\_C\_\lambda-1.3}$ & 0.0 & 560 & 0.01 & 10.4 & 9.9 & 0.04 & -1.30 & 0.55 & 11.9 & 0.53 & 28 & 490 & 3400\\
$\mathtt{Ms9.1\_C\_\lambda-0.3}$ & 0.0 & 400 & 0.03 & 9.1 & 7.6 & 0.04 & -0.33 & 0.61 & 11.2 & 0.61 & 31 & 500 & 3100\\
$\mathtt{Ms9.2\_C\_\lambda+0.9}$ & 0.0 & 400 & 0.10 & 9.2 & 6.8 & 0.04 & 0.86 & 0.65 & 9.4 & 0.62 & 33 & 630 & 3800\\
$\mathtt{Ms9.3\_C\_\lambda+2.0}$ & 0.0 & 400 & 0.30 & 9.3 & 5.8 & 0.04 & 1.98 & 0.67 & 9.7 & 0.73 & 33 & 630 & 3700\\
$\mathtt{Ms10.6\_C\_\lambda+3.3}$ & 0.0 & 400 & 1.00 & 10.6 & 4.0 & 0.04 & 3.27 & 0.65 & 8.6 & 0.81 & 37 & 910 & 5000\\
$\mathtt{Ms10.1\_C\_\lambda+4.3}$ & 0.0 & 400 & 3.00 & 10.1 & 3.7 & 0.04 & 4.34 & 0.65 & 8.9 & 0.88 & 36 & 800 & 4500\\
[0.3em]
\bottomrule\end{tabular}
  \caption{Parameters of our compressively-driven simulations. Columns show the run name, solenoidal weight, $\zeta$, acceleration, $a$, forcing correlation time, $\tau_\mathrm{a}$, volume-weighted RMS Mach number, $M_\mathrm{V}$, mass-weighted RMS Mach number, $M_\mathrm{M}$, eddy turnover time, $\tau_\mathrm{e}$, relative forcing correlation time, $\lambda_{\rm a} \equiv {\rm ln}(\tau_\mathrm{a} / \tau_\mathrm{e}$), small-scale compressive ratio, $r_\mathrm{cs}$, effective kinematic viscosity $\nu_{\rm eff}$, effective Kolmogorov scale $\eta$, Taylor microscale, $\lambda$, and the Taylor microscale and integral scale Reynolds numbers,  $\mathrm{Re}_{\lambda}$ and $\mathrm{Re}_{\rm int}$.   The standard deviation of all simulation properties in each simulation is below $10\%$   except for $\nu_{\rm eff}$ (and derived properties) with a maximum of $\approx50\%$ -- especially in the  $\tau_\mathrm{a}=3$ cases. For all dimensional quantities, the unit of length is the box size and the unit of time is the box sound crossing time. All simulations were carried out on a fixed grid of $512^3$ cells with a fixed viscosity of $\nu = 5.5 \times 10^{-4}$. The detailed definitions of the quantities are given in Sec.~\ref{sec:parameters}.  }
\label{tab:runsC}
\end{table*}%

\begin{table*}[t]\centering\begin{tabular}{lccccccccccccc}\toprule
& \multicolumn{3}{c}{Input parameters} & \multicolumn{10}{c}{Simulation properties} \\
\cmidrule(lr{.75em}){2-4} \cmidrule(lr{.75em}){5-14}
\texttt{Id} & $\zeta$ & $a$ & $\tau_\mathrm{a}$ & $M_\mathrm{V}$ & $M_\mathrm{M}$ & $\tau_\mathrm{e}$ & ln$(\tau_\mathrm{a} / \tau_\mathrm{e})$ & $r_\mathrm{cs}$ & $\nu_{\rm eff}~[10^{-4}]$ & $\eta/\Delta_x$ & $\lambda/\Delta_x$ & $\mathrm{Re}_{\lambda}$ & $\mathrm{Re}_{\rm int}$ \\ \midrule
$\mathtt{Ms2.0\_M\_\lambda-3.0}$ & 0.3 & 35 & 0.01 & 2.0 & 1.9 & 0.20 & -3.01 & 0.77 & 8.0 & 1.31 & 50 & 240 & 1000\\
$\mathtt{Ms2.1\_M\_\lambda-1.8}$ & 0.3 & 25 & 0.03 & 2.1 & 2.0 & 0.19 & -1.85 & 0.75 & 7.9 & 1.27 & 48 & 250 & 1100\\
$\mathtt{Ms2.4\_M\_\lambda-0.5}$ & 0.3 & 25 & 0.10 & 2.4 & 2.2 & 0.17 & -0.52 & 0.72 & 7.8 & 1.20 & 45 & 270 & 1200\\
$\mathtt{Ms2.5\_M\_\lambda+0.6}$ & 0.3 & 25 & 0.30 & 2.5 & 2.1 & 0.16 & 0.60 & 0.67 & 7.4 & 1.20 & 43 & 280 & 1400\\
$\mathtt{Ms2.5\_M\_\lambda+1.8}$ & 0.3 & 25 & 1.00 & 2.5 & 2.0 & 0.16 & 1.83 & 0.64 & 7.3 & 1.24 & 41 & 270 & 1400\\
$\mathtt{Ms2.6\_M\_\lambda+3.0}$ & 0.3 & 25 & 3.00 & 2.6 & 2.3 & 0.16 & 2.95 & 0.62 & 7.3 & 1.17 & 42 & 300 & 1500\\
[0.3em]
$\mathtt{Ms3.5\_M\_\lambda-0.2}$ & 0.3 & 50 & 0.10 & 3.5 & 3.0 & 0.12 & -0.15 & 0.70 & 8.2 & 0.98 & 41 & 340 & 1700\\
[0.3em]
$\mathtt{Ms5.0\_M\_\lambda-2.1}$ & 0.3 & 140 & 0.01 & 5.0 & 4.7 & 0.08 & -2.11 & 0.67 & 9.2 & 0.77 & 38 & 410 & 2200\\
$\mathtt{Ms4.8\_M\_\lambda-1.0}$ & 0.3 & 100 & 0.03 & 4.8 & 4.3 & 0.08 & -1.04 & 0.67 & 9.0 & 0.80 & 39 & 400 & 2200\\
$\mathtt{Ms5.0\_M\_\lambda+0.2}$ & 0.3 & 100 & 0.10 & 5.0 & 4.1 & 0.08 & 0.21 & 0.66 & 8.6 & 0.80 & 38 & 430 & 2400\\
$\mathtt{Ms5.1\_M\_\lambda+1.3}$ & 0.3 & 100 & 0.30 & 5.1 & 3.7 & 0.08 & 1.33 & 0.66 & 8.5 & 0.84 & 38 & 450 & 2500\\
$\mathtt{Ms5.2\_M\_\lambda+2.6}$ & 0.3 & 100 & 1.00 & 5.2 & 3.9 & 0.08 & 2.56 & 0.64 & 8.0 & 0.81 & 38 & 480 & 2600\\
$\mathtt{Ms5.4\_M\_\lambda+3.7}$ & 0.3 & 100 & 3.00 & 5.4 & 3.9 & 0.08 & 3.66 & 0.62 & 8.1 & 0.83 & 39 & 520 & 2800\\
[0.3em]
$\mathtt{Ms7.3\_M\_\lambda+0.6}$ & 0.3 & 200 & 0.10 & 7.3 & 5.8 & 0.06 & 0.59 & 0.61 & 9.2 & 0.67 & 34 & 530 & 3200\\
[0.3em]
$\mathtt{Ms10.9\_M\_\lambda-1.3}$ & 0.3 & 560 & 0.01 & 10.9 & 10.4 & 0.04 & -1.28 & 0.50 & 11.7 & 0.51 & 28 & 520 & 3700\\
$\mathtt{Ms10.1\_M\_\lambda-0.3}$ & 0.3 & 400 & 0.03 & 10.1 & 9.5 & 0.04 & -0.27 & 0.52 & 11.0 & 0.54 & 29 & 540 & 3700\\
$\mathtt{Ms10.9\_M\_\lambda+1.0}$ & 0.3 & 400 & 0.10 & 10.9 & 8.5 & 0.04 & 0.99 & 0.54 & 10.6 & 0.56 & 31 & 630 & 4200\\
$\mathtt{Ms10.7\_M\_\lambda+2.1}$ & 0.3 & 400 & 0.30 & 10.7 & 8.0 & 0.04 & 2.08 & 0.54 & 10.0 & 0.58 & 30 & 640 & 4400\\
$\mathtt{Ms11.4\_M\_\lambda+3.3}$ & 0.3 & 400 & 1.00 & 11.4 & 7.6 & 0.04 & 3.31 & 0.53 & 9.7 & 0.57 & 32 & 760 & 5000\\
$\mathtt{Ms11.2\_M\_\lambda+4.4}$ & 0.3 & 400 & 3.00 & 11.2 & 8.2 & 0.04 & 4.42 & 0.53 & 10.0 & 0.56 & 31 & 690 & 4600\\
[0.7em]
$\mathtt{Ms3.0\_S\_\lambda-0.3}$ & 1.0 & 25 & 0.10 & 3.0 & 2.8 & 0.13 & -0.29 & 0.62 & 7.8 & 1.05 & 41 & 310 & 1600\\
[0.3em]
$\mathtt{Ms4.4\_S\_\lambda+0.1}$ & 1.0 & 50 & 0.10 & 4.4 & 4.1 & 0.09 & 0.07 & 0.60 & 8.6 & 0.84 & 38 & 380 & 2100\\
[0.3em]
$\mathtt{Ms5.3\_S\_\lambda-2.0}$ & 1.0 & 140 & 0.01 & 5.3 & 5.0 & 0.08 & -2.02 & 0.58 & 9.0 & 0.73 & 35 & 400 & 2300\\
$\mathtt{Ms5.5\_S\_\lambda-0.9}$ & 1.0 & 100 & 0.03 & 5.5 & 5.2 & 0.07 & -0.90 & 0.57 & 9.0 & 0.72 & 35 & 420 & 2400\\
$\mathtt{Ms6.2\_S\_\lambda+0.4}$ & 1.0 & 100 & 0.10 & 6.2 & 5.9 & 0.07 & 0.42 & 0.54 & 9.0 & 0.66 & 34 & 460 & 2900\\
$\mathtt{Ms6.5\_S\_\lambda+1.5}$ & 1.0 & 100 & 0.30 & 6.5 & 6.2 & 0.06 & 1.55 & 0.53 & 9.0 & 0.65 & 34 & 480 & 3100\\
$\mathtt{Ms6.9\_S\_\lambda+2.8}$ & 1.0 & 100 & 1.00 & 6.9 & 6.4 & 0.06 & 2.78 & 0.51 & 8.9 & 0.63 & 34 & 520 & 3400\\
$\mathtt{Ms7.0\_S\_\lambda+3.9}$ & 1.0 & 100 & 3.00 & 7.0 & 6.9 & 0.06 & 3.87 & 0.49 & 8.9 & 0.62 & 35 & 580 & 3700\\
[0.3em]
$\mathtt{Ms8.9\_S\_\lambda+0.8}$ & 1.0 & 200 & 0.10 & 8.9 & 8.5 & 0.05 & 0.79 & 0.47 & 9.7 & 0.53 & 30 & 540 & 3800\\
[0.3em]
$\mathtt{Ms12.3\_S\_\lambda+1.2}$ & 1.0 & 400 & 0.10 & 12.3 & 12.2 & 0.03 & 1.09 & 0.40 & 11.7 & 0.47 & 27 & 590 & 4700\\
  \bottomrule\end{tabular}
\caption{Parameters of our mixed and solenoidally-driven simulations. Columns are as in Table \ref{tab:runsC}.}
\label{tab:runsMS}
\end{table*}

\begin{figure*}[t]
\begin{center}
\includegraphics[width=1.0\textwidth]{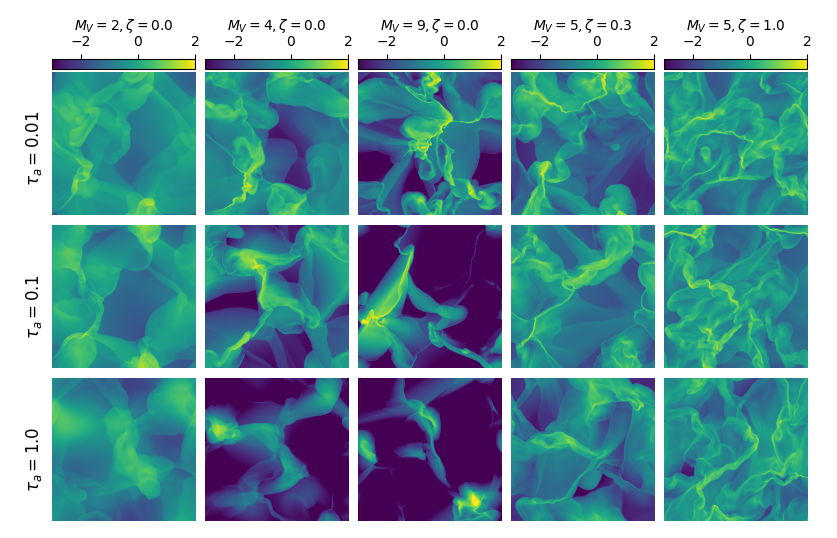}
\end{center}
\vspace{-0.2in}
\caption{Representative results from our  simulations. From left to right, columns show results slices of $s$ from runs with compressive driving, and Mach numbers $M_{\rm V} \approx$ 2, 4, and 8, mixed-driving runs with $M_{\rm V} \approx 4$ runs, and purely solenoidal runs with $M_{\rm V} \approx 4.$ From top to bottom, the rows show cases with $\tau_{\rm a}$ = 0.01, 0.1, and 1. The corresponding run names are (top left to bottom right): $\mathtt{Ms1.9\_C\_\lambda-3.1}$,
$\mathtt{Ms4.7\_C\_\lambda-2.1}$,
$\mathtt{Ms10.4\_C\_\lambda-1.3}$,
$\mathtt{Ms5.0\_M\_\lambda-2.1}$,
$\mathtt{Ms5.3\_S\_\lambda-2.0}$,
$\mathtt{Ms2.0\_C\_\lambda-0.7}$,
$\mathtt{Ms4.3\_C\_\lambda-0.1}$,
$\mathtt{Ms9.2\_C\_\lambda-0.9}$,
$\mathtt{Ms5.0\_M\_\lambda-0.2}$,
$\mathtt{Ms6.2\_S\_\lambda-0.4}$,
$\mathtt{Ms1.4\_C\_\lambda-1.4}$,
$\mathtt{Ms4.0\_C\_\lambda-2.4}$,
$\mathtt{Ms10.6\_C\_\lambda-3.3}$,
$\mathtt{Ms5.2\_M\_\lambda-2.6}$,
$\mathtt{Ms6.9\_S\_\lambda-2.8}$. Snapshots are shown at $t \approx 8 \tau_e$. At that time, the results are stable and do not vary.}
\label{fig:slices}
\end{figure*}

We carried out the simulations with the \texttt{AthenaPK} code,\footnote{ \texttt{AthenaPK} is available and maintained at \url{https://github.com/parthenon-hpc-lab/athenapk} and commit \texttt{80942e8} was used for the simulations.} which implements finite-volume hydrodynamic and magnetohydrodynamic algorithms on the \texttt{Parthenon} adaptive mesh refinement (AMR) framework \citep{parthenon}. This framework, derived from \texttt{Athena++} \citep{Stone20}, \texttt{K-Athena} \citep{kathena}, and \texttt{Kokkos} \citep{Trott21}, provides excellent computational efficiency and scalability across various GPU architectures. 

Our numerical setup employed an unsplit second-order finite-volume method with a predictor-corrector time integration scheme (Van Leer), an HLLC Riemann solver, and piecewise parabolic reconstruction in primitive variables. To maintain near-isothermal conditions, we used an ideal equation of state with an adiabatic index \( \gamma = 1.0001 \). We computed the viscous fluxes at cell interfaces using second-order finite differences and integrated them alongside Riemann fluxes in an unsplit manner. In cases where higher-order updates produced unphysically negative densities or pressures, we applied a first-order flux correction, reverting to piecewise-constant reconstruction and an LLF Riemann solver. 

To drive turbulence, we employed a stochastic forcing mechanism governed by an Ornstein-Uhlenbeck equation \citep{Schmidt09, Grete2018}. 
In Fourier space, this can be expressed as
\be
\hat a_i (\mathbf{k},t + \Delta t) = c_\mathrm{drift} \hat a_i (\mathbf{k},t) + \sqrt{1-c_\mathrm{drift}^2} P_{\rm a}(k) \mathcal{P}_{ij} \mathcal{N}_j.
\label{eq:driving}
\ee
Here, $c_\mathrm{drift}=e^{-\Delta t/\tau_\mathrm{a}}$ is the drift coefficient and $\sqrt{1-c_\mathrm{drift}^2}$ the diffusion coefficient, i.e., $\tau_\mathrm{a}$ sets the correlation time of the driving, which we vary as described in detail below.
\( P_{\rm a}(k) \) defines the spectral profile of the acceleration field, which peaks at \( k_p =2\)  (i.e., half the box size).  It has
 the form 
\begin{equation}
P_{\rm a}(k) = \tilde k^2 (2 - \tilde k^2) \Theta(\tilde k^2 - 2),
\end{equation}
where \( \Theta \) is the Heaviside step function and \( \tilde k \equiv k/k_p \), and
  $\mathcal{N}_j$ are complex random numbers with $0 < |\mathcal{N}_j|<1$ and zero mean, for which the real and imaginary parts are independently drawn from a uniform distribution. 
  The projection tensor
 \( \mathcal{P}_{ij} = \left[ \zeta \delta_{ij} + (1-2\zeta) \frac{k_i k_j}{|\mathbf{k}|^2} \right] \) determines the partitioning of the driving energy between solenoidal and compressive modes via a Helmholtz decomposition. The parameter \( \zeta \in [0,1] \) regulates this partitioning: \( \zeta =0 \) corresponds to purely compressive driving, while \( \zeta = 1 \) results in purely solenoidal forcing. 

\subsection{Parameter Space}
\label{sec:parameters}

We conducted our simulations on a fixed grid with $512^3$ cells. Following \cite{Paper2}, all cases included an explicit viscosity of $5.5 \times 10^{-4}$ in units of the box size and sound speed. However, the simulations should still be considered implicit large eddy simulations and not direct numerical simulations because the effective viscosity due to the numerical scheme is about twice as large as the explicit one.

Tables~\ref{tab:runsC} and \ref{tab:runsMS} list the key parameters of our simulations, each of which is named after the Mach number, the nature of the driving, and the log of the ratio of the driving and eddy turnover times. The simulations are split into three sets. The largest set consists of 30 fully compressive ($\zeta=0$) runs, with the strength and correlation time of the driving accelerations chosen to regularly span a wide range of Mach numbers and $\lambda_{\rm a} \equiv \ln(\tau_{\rm a}/\tau_{\rm e})$ values. These simulations are divided into 5 groups of 6, with the driving strength chosen such that the simulations within each group have roughly the same Mach number ($\approx$ 2, 3, 4, 6, and 9).  In the tables, the simulations are sorted by Mach number group, and within each group, they are sorted by increasing values of the driving correlation time. Throughout this paper, all times are in units of the sound crossing time, $\tau_{\rm sc}$.

Tables~\ref{tab:runsC} and \ref{tab:runsMS} also give the volume-weighted and mass-weighted Mach numbers ($M_{\rm V}$ and $M_{\rm M} $) of each run. These are computed as the rms average within the stationary regime, defined approximately as the period between $\max(5 \tau_{\rm e}, \tau_{\rm a})$ and  $\max(10 \tau_{\rm e}, 2 \tau_{\rm a}).$  Here, the eddy turnover time is computed as $\tau_{\rm e} \equiv L_i/M_{\rm V}$ using the integral scale, defined as $L_i \equiv \int E(k)/k \, \mathrm{d}k / \int E(k) \, \mathrm{d}k \approx 0.38$ where $E(k)$ is the specific kinetic energy spectrum. 

For each group of compressively-driven runs, we vary $\tau_{\rm a}$ from 0.01 to 3 sound crossing times, logarithmically sampling a large range of $\tau_{\rm a}/\tau_{\rm e}$ values. In the mixed-driving, $\zeta=0.3,$ case, we carried out three groups of 6 simulations that span $\tau_{\rm a}$ from 0.01 to 3 for Mach numbers of $M_{\rm V} \approx 2, 5,$ and 10, and two simulations that probe $M_{\rm V} \approx 3.5$ and 7 for a fixed $\tau_{\rm a}$ of 0.1.  Finally, in the solenoidal driving, $\zeta=1.0,$ case we carried out runs with $M_{\rm V} \approx 3, 4, 8,$ and 12 cases with $\tau_{\rm a}$ of 0.1 and a single group of simulations that span $\tau_{\rm a}$ from 0.01 to 3 with $M_{\rm V} \approx 5.$ 

Tables~\ref{tab:runsC} and~\ref{tab:runsMS} also provide numerical values for the small-scale compressive ratio \citep{Kida1990}, $r_\mathrm{cs} = \left < |\nabla \cdot \mathbf{u}|^2 \right >_{\rm V} / (\left < |\nabla \cdot \mathbf{u}|^2 \right >_{\rm V} + \left < |\nabla \times \mathbf{u}|^2 \right >_{\rm V})$, the effective viscosity \citep{Paper2}, $\nu_\mathrm{eff}$, the resulting effective Kolmogorov scale, $\eta = \left ( \nu_{\rm eff}^3 / \dot E_{\rm e} \right )^{1/4}$, and effective integral scale Reynolds numbers $\mathrm{Re_{\rm int}} = M_{\rm V} c_s L_i / \nu_{\rm eff}$. Note that here  we use the rms velocity (rather than the mean of the velocity fluctuations, which differs by $1/\sqrt{3}$ in isotropic turbulence) as commonly done in the astrophysical literature.

Similarly, as in \citet{Paper2}, we calculate the Taylor microscale,$\lambda = \sqrt{5 \left < |\mathbf{u}|^2 \right >_{\rm V}/ \left < |\nabla \times \mathbf{u}|^2 \right >_{\rm V}}$, and associated Reynolds number, $\mathrm{Re}_\lambda = M_{\rm V} c_s \lambda / \nu_{\rm eff}$. In our simulations, the Kolmogorov scale is of the order of the grid scale, and the shock width is also about the cell size, as artificially set by the numerical code. The numerical resolution is sufficient to resolve the postshock density, and properly capture the low-order statistics of density fluctuations we are focusing on in this study \citep[e.g.][]{Yamazaki2002}.  The Taylor microscale is $\approx 4 - 7 \times$ smaller than the integral lengthscale, indicating the presence of a (limited) inertial range, as also expected from Reynolds numbers   $\mathrm{Re_{\rm int}} > 1000$. In terms of these turbulence characteristics, the simulations occupy a largely comparable region of parameter space. Their key differences stem from the driving parameters, which in turn determine the Mach numbers and density statistics.

While the dynamical range in our simulations is limited, our previous work has shown that $512^3$ resolution is sufficient to accurately reproduce the PDF of $s$ at similar Mach numbers over a range of driving mechanisms and correlation times \citep{Paper2}. Similarly, \citet{Konstandin12} show that for increased dynamical range, the trends of the density PDF up to at least $M \approx 5.5$ are be independent of resolution (see Sec. 3.2 and Fig 4 of this work). As a further test of convergence, we restarted the simulation at the most extreme point in parameter space in our suite, Ms10.1\_C\_$\lambda$4.3 at $t = 10 \tau_{\rm e}$ and ran it at $1024^3$ resolution and lower viscosity ($3.5 \times 10^{-4}$) for another eddy turnover time. Although the small-scale structure was sharper, $P_{\rm V}(s)$ was indistinguishable from that of our fiducial runs.

\section{Results}

\begin{figure*}[t]
\begin{center}
\includegraphics[width=\textwidth]{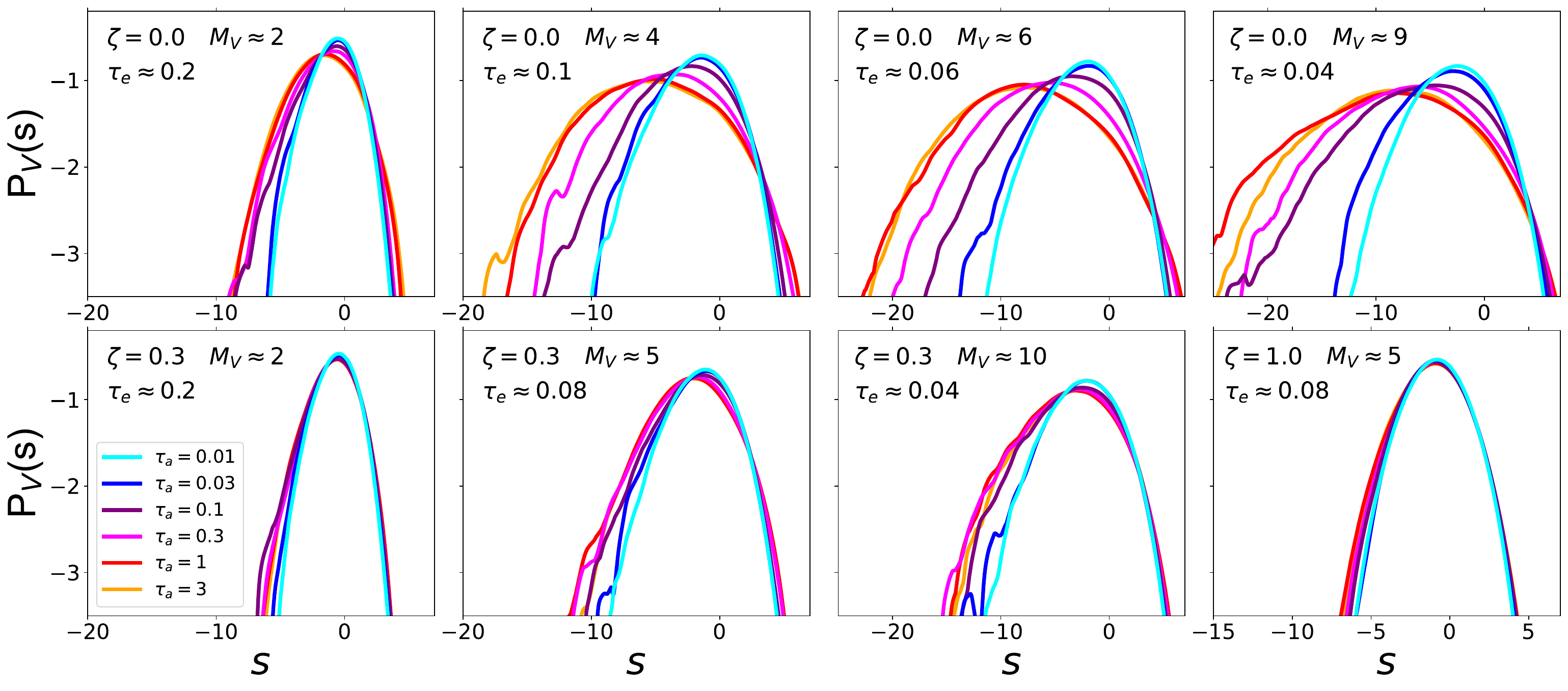}
\end{center}
\vspace{-0.2in}
\caption{Volume-weighted PDFs from a representative subset of our simulations. Here, the top row shows the results of simulations with fully compressive driving and average volume-weighted Mach numbers of $M_{\rm V} \approx 2,4,6,$ and 9. The lower row shows simulation results with mixed driving, and $M_{\rm V} \approx 2,5,$ and 10, as well as results from solenoidally-driven simulations with $M_{\rm V} \approx 5.$ In each panel, the colored lines show $P_{\rm V}(s)$ for runs with $\tau_{\rm a}$ = 0.01 (cyan), 0.03 (blue), 0.1 (purple), 0.3 (magenta), 1.0 (red), and 3.0 (orange). Increasing $\tau_{\rm a}$ has a strong effect on the compressive runs, broadening $P_{\rm V}(s)$ and moving the peak to the left, consistent with the formation of large voids. These effects are seen to a limited degree in the mixed driving runs, while the differences in the solenoidal results are consistent with those expected due to the small differences in $M_{\rm V}$ between the various runs.}
\label{fig:PSV}
\end{figure*}

\begin{figure*}[t]
\begin{center}
\includegraphics[width=\textwidth]{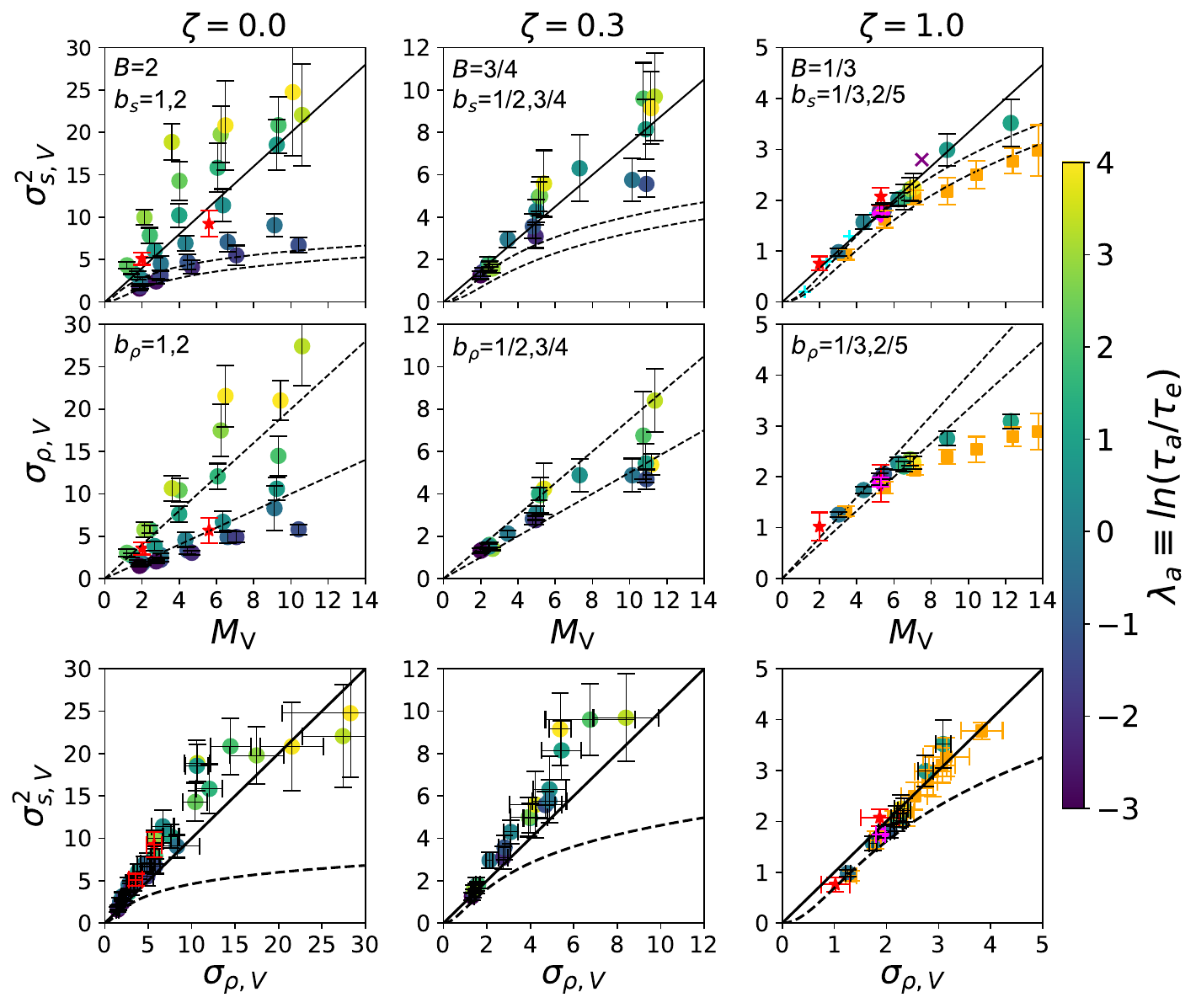}
\end{center}
\vspace{-0.3in}
\caption{Volume-weighted variance of $s,$ $\sigma^2_{\rm s,V}$ (top row), and standard deviation of  $\rho/\rho_0,$ $\sigma_{\rho,V}$ (center row), as a function of Mach number, and $\sigma^2_{\rm s,V}$ as a function of $\sigma_{\rho,V}$ (bottom row). In all rows, the filled circles are the results of our simulations, with the colors corresponding to the $\lambda_{\rm a}$ values.  The other points are taken from \citet[][cyan crosses]{Lemaster08}, \citet[][magenta diamonds]{Federrath10}, \citet[orange squares]{Price11}, \citet[][red stars]{Konstandin12}, and \cite[][purple x]{Pan18}.  In the upper row, the dashed lines show fits of the form $\sigma^2_{s,{\rm V}} = \ln(1+M_{\rm V}^2 b_s^2 )$, and the solid lines show fits of the form $\sigma^2_{s,{\rm V}} = B \, M_{\rm V}$ with $b_s$ and $B$ values labeled in each panel.  In the center row, the dashed lines show fits of the form $\sigma_{\rho,{\rm V}} = b_{\rho} M_{\rm V}$. In the bottom row, the dashed lines show $\sigma^2_{s,{\rm V}}= \ln(1+\sigma_{\rho,{\rm V}}^2)$ as expected for a Gaussian distribution of $s$ and the solid lines show $\sigma^2_{s,{\rm V}}= \sigma_{\rho,{\rm V}}.$ These panels show that the primary problem with eq.\ (\ref{eq:standardfit}) is that $P_{\rm V}(s)$ cannot be adequately approximated by a Gaussian.}
\label{fig:sigrho_sig2s}
\end{figure*}

\subsection{Spatial Distribution}

Figure~\ref{fig:slices} gives a visual representation of the results of our simulations, contrasting runs with $\tau_{\rm a}=0.01,$ 0.1, and 1.0 for several key values of $\zeta$ and Mach number. Here, we see that, in the compressively-driven simulations, the runs with longer correlation times contain large voids, whose prominence increases with increasing $\tau_{\rm a}.$ As discussed in \cite{Paper2}, these are regions in which $\nabla \cdot \mathbf a$ is positive and accelerated expansions are sustained over a significant period of time. This results in large, expanding regions, ringed by shocks of swept-up material.

One consequence of these expanding voids is that underdense regions tend to have larger velocities than denser regions. This means that the ratio of mass-weighted and volume-weighted Mach numbers $M_{\rm M}/M_{\rm V}$ decreases from $\approx 1$ to $\lesssim  0.5$ as $\lambda_{\rm a}$ increases from $\approx -3$ to $\approx 3$, as can be seen in Table \ref{tab:runsC}.

As these voids are a direct result of $\nabla \cdot \mathbf a,$ their impact becomes weaker in the mixed driving $\zeta=0.3$ simulations, which display only minor changes at large $\tau_{\rm a}$ values. Similarly, while $M_{\rm M}/M_{\rm V}$ generally decreases in these runs as $\tau_{\rm a}$ increases, the effect is much less than in the compressive case.

 Finally, in the solenoidal case, $\nabla \cdot \mathbf a= 0.$ This means that the driving force does not directly influence the divergence $\nabla \cdot {\mathbf v},$ and hence does not directly change the density distribution. Instead, the accelerations cause compressions and expansions only indirectly through nonlinear interactions, which lead to the formation of shocks.  Thus, the distribution of $s$ shows no detectable dependence on $\tau_{\rm a},$ and, likewise, $\tau_{\rm a}$ has little effect on the relation between the mass and volume-weighted Mach numbers.

\begin{table*}[t]
\hspace{-0.8in}
 \centering
 \resizebox{1.1\textwidth}{!}{
\begin{tabular}{lccccccccccc}
\toprule
& \multicolumn{3}{c}{Flow Properties} & \multicolumn{4}{c}{PDF Moments} & \multicolumn{4}{c}{Derived Values}\\
\cmidrule(lr{.75em}){2-4} \cmidrule(lr{.75em}){5-8} \cmidrule(lr{.75em}){9-12}
Name & $M_{\rm V}$ & $M_{\rm M}$ & $\lambda_{\rm a}$ &
$\left<s\right>_{\rm V}$ & $\sigma_{s,{\rm V}}^2$ & $\mu_{s,{\rm V}}$& $\sigma_{\rho,{\rm V}}$ & $b_s$ & $b_{\rho}$ &
$B$ & $\left<s\right>_{\rm V}$ $M_{\rm V}^{-1}$ \\

$\mathtt{Ms1.9\_C\_\lambda-3.1}$ & 1.9 & 1.8 & -3.07 & -0.72$\pm$0.09 & 1.59$\pm$0.27 & -0.23$\pm$0.22 & 1.49$\pm$0.14 & 1.06 & 0.80$\pm$0.08 & 0.85$\pm$0.15 & -0.39$\pm$0.05 \\
$\mathtt{Ms1.9\_C\_\lambda-1.9}$ & 1.9 & 1.8 & -1.94 & -0.80$\pm$0.08 & 1.83$\pm$0.27 & -0.30$\pm$0.23 & 1.58$\pm$0.13 & 1.19 & 0.82$\pm$0.07 & 0.95$\pm$0.15 & -0.42$\pm$0.05 \\
$\mathtt{Ms2.0\_C\_\lambda-0.7}$ & 2.0 & 1.8 & -0.67 & -1.04$\pm$0.11 & 2.58$\pm$0.41 & -0.42$\pm$0.25 & 1.78$\pm$0.17 & 1.73 & 0.88$\pm$0.08 & 1.28$\pm$0.21 & -0.51$\pm$0.06 \\
$\mathtt{Ms1.8\_C\_\lambda+0.3}$ & 1.8 & 1.4 & 0.34 & -1.27$\pm$0.13 & 3.14$\pm$0.50 & -0.32$\pm$0.22 & 2.01$\pm$0.15 & 2.66 & 1.14$\pm$0.09 & 1.77$\pm$0.29 & -0.72$\pm$0.08 \\
$\mathtt{Ms1.4\_C\_\lambda+1.4}$ & 1.4 & 1.0 & 1.36 & -1.58$\pm$0.11 & 3.60$\pm$0.31 & -0.14$\pm$0.13 & 2.75$\pm$0.24 & 4.17 & 1.92$\pm$0.16 & 2.52$\pm$0.23 & -1.10$\pm$0.08 \\
$\mathtt{Ms1.2\_C\_\lambda+2.3}$ & 1.2 & 0.7 & 2.28 & -1.86$\pm$0.13 & 4.33$\pm$0.40 & -0.11$\pm$0.12 & 3.04$\pm$0.43 & 7.27 & 2.54$\pm$0.36 & 3.64$\pm$0.42 & -1.56$\pm$0.15 \\[0.3em]
$\mathtt{Ms2.8\_C\_\lambda-2.7}$ & 2.8 & 2.6 & -2.69 & -1.06$\pm$0.10 & 2.43$\pm$0.32 & -0.22$\pm$0.21 & 2.03$\pm$0.19 & 1.17 & 0.73$\pm$0.07 & 0.88$\pm$0.13 & -0.38$\pm$0.04 \\
$\mathtt{Ms3.0\_C\_\lambda-1.5}$ & 3.0 & 2.8 & -1.49 & -1.28$\pm$0.12 & 3.16$\pm$0.56 & -0.35$\pm$0.25 & 2.22$\pm$0.19 & 1.58 & 0.74$\pm$0.06 & 1.05$\pm$0.19 & -0.43$\pm$0.04 \\
$\mathtt{Ms3.0\_C\_\lambda-0.3}$ & 3.0 & 2.5 & -0.27 & -1.73$\pm$0.19 & 4.52$\pm$0.89 & -0.35$\pm$0.23 & 2.74$\pm$0.25 & 3.16 & 0.91$\pm$0.08 & 1.50$\pm$0.30 & -0.57$\pm$0.07 \\
$\mathtt{Ms2.7\_C\_\lambda+0.8}$ & 2.7 & 1.9 & 0.76 & -2.37$\pm$0.23 & 6.08$\pm$0.87 & -0.24$\pm$0.19 & 3.83$\pm$0.53 & 7.79 & 1.43$\pm$0.20 & 2.27$\pm$0.34 & -0.88$\pm$0.09 \\
$\mathtt{Ms2.4\_C\_\lambda+1.9}$ & 2.4 & 1.3 & 1.88 & -3.32$\pm$0.22 & 7.86$\pm$0.90 & -0.07$\pm$0.13 & 5.79$\pm$0.45 & 21.0 & 2.39$\pm$0.19 & 3.25$\pm$0.47 & -1.37$\pm$0.15 \\
$\mathtt{Ms2.1\_C\_\lambda+2.9}$ & 2.1 & 1.1 & 2.89 & -3.95$\pm$0.23 & 9.98$\pm$0.92 & -0.00$\pm$0.11 & 5.76$\pm$0.90 & 68.9 & 2.70$\pm$0.42 & 4.69$\pm$0.57 & -1.85$\pm$0.18 \\[0.3em]
$\mathtt{Ms4.7\_C\_\lambda-2.1}$ & 4.7 & 4.4 & -2.15 & -1.70$\pm$0.15 & 4.12$\pm$0.82 & -0.23$\pm$0.30 & 3.04$\pm$0.28 & 1.66 & 0.65$\pm$0.06 & 0.88$\pm$0.18 & -0.36$\pm$0.04 \\
$\mathtt{Ms4.4\_C\_\lambda-1.1}$ & 4.4 & 3.9 & -1.10 & -1.91$\pm$0.24 & 4.72$\pm$0.91 & -0.25$\pm$0.26 & 3.32$\pm$0.44 & 2.37 & 0.75$\pm$0.10 & 1.06$\pm$0.21 & -0.43$\pm$0.06 \\
$\mathtt{Ms4.3\_C\_\lambda+0.1}$ & 4.3 & 3.5 & 0.11 & -2.67$\pm$0.29 & 6.94$\pm$0.90 & -0.29$\pm$0.28 & 4.57$\pm$0.85 & 7.42 & 1.06$\pm$0.20 & 1.60$\pm$0.23 & -0.62$\pm$0.08 \\
$\mathtt{Ms4.0\_C\_\lambda+1.2}$ & 4.0 & 2.5 & 1.15 & -4.03$\pm$0.26 & 10.23$\pm$1.39 & -0.10$\pm$0.22 & 7.62$\pm$0.92 & 41.6 & 1.91$\pm$0.23 & 2.56$\pm$0.39 & -1.01$\pm$0.09 \\
$\mathtt{Ms4.0\_C\_\lambda+2.3}$ & 4.0 & 1.7 & 2.38 & -5.87$\pm$0.55 & 14.29$\pm$2.27 & 0.07$\pm$0.13 & 10.42$\pm$1.40 & 316 & 2.60$\pm$0.35 & 3.56$\pm$0.65 & -1.46$\pm$0.19 \\
$\mathtt{Ms3.6\_C\_\lambda+3.4}$ & 3.6 & 1.3 & 3.39 & -7.21$\pm$0.55 & 18.88$\pm$2.16 & 0.18$\pm$0.11 & 10.65$\pm$1.48 & 3490 & 2.96$\pm$0.41 & 5.24$\pm$0.74 & -2.00$\pm$0.23 \\[0.3em]
$\mathtt{Ms7.1\_C\_\lambda-1.7}$ & 7.1 & 6.9 & -1.72 & -2.35$\pm$0.28 & 5.51$\pm$1.16 & -0.14$\pm$0.25 & 4.90$\pm$0.65 & 2.22 & 0.69$\pm$0.09 & 0.78$\pm$0.17 & -0.33$\pm$0.05 \\
$\mathtt{Ms6.6\_C\_\lambda-0.7}$ & 6.6 & 5.7 & -0.70 & -2.61$\pm$0.20 & 7.07$\pm$1.16 & -0.36$\pm$0.22 & 4.90$\pm$0.76 & 5.20 & 0.74$\pm$0.11 & 1.07$\pm$0.18 & -0.40$\pm$0.03 \\
$\mathtt{Ms6.3\_C\_\lambda+0.5}$ & 6.3 & 4.8 & 0.48 & -3.88$\pm$0.36 & 11.41$\pm$1.92 & -0.30$\pm$0.23 & 6.69$\pm$1.27 & 47.4 & 1.05$\pm$0.20 & 1.80$\pm$0.32 & -0.61$\pm$0.06 \\
$\mathtt{Ms6.1\_C\_\lambda+1.6}$ & 6.1 & 3.7 & 1.57 & -5.83$\pm$0.58 & 15.84$\pm$2.90 & -0.06$\pm$0.11 & 12.05$\pm$1.51 & 454 & 1.99$\pm$0.25 & 2.61$\pm$0.54 & -0.96$\pm$0.13 \\
$\mathtt{Ms6.2\_C\_\lambda+2.8}$ & 6.2 & 2.8 & 2.79 & -7.66$\pm$0.58 & 19.78$\pm$3.34 & 0.07$\pm$0.18 & 17.48$\pm$3.10 & 3162 & 2.81$\pm$0.50 & 3.17$\pm$0.60 & -1.23$\pm$0.14 \\
$\mathtt{Ms6.5\_C\_\lambda+3.9}$ & 6.5 & 1.8 & 3.94 & -8.92$\pm$0.98 & 20.84$\pm$5.24 & 0.13$\pm$0.22 & 21.54$\pm$3.62 & 5170 & 3.32$\pm$0.56 & 3.22$\pm$0.95 & -1.38$\pm$0.26 \\[0.3em]
$\mathtt{Ms10.4\_C\_\lambda-1.3}$ & 10.4 & 9.9 & -1.30 & -2.79$\pm$0.18 & 6.71$\pm$0.88 & -0.13$\pm$0.18 & 5.80$\pm$0.60 & 2.75 & 0.56$\pm$0.06 & 0.64$\pm$0.09 & -0.27$\pm$0.02 \\
$\mathtt{Ms9.1\_C\_\lambda-0.3}$ & 9.1 & 7.7 & -0.33 & -3.45$\pm$0.24 & 9.06$\pm$1.34 & -0.22$\pm$0.19 & 8.30$\pm$2.60 & 10.1 & 0.91$\pm$0.29 & 0.99$\pm$0.15 & -0.38$\pm$0.03 \\
$\mathtt{Ms9.2\_C\_\lambda+0.9}$ & 9.2 & 6.8 & 0.86 & -5.65$\pm$0.52 & 18.55$\pm$2.98 & -0.32$\pm$0.26 & 10.60$\pm$1.33 & 1150 & 1.15$\pm$0.14 & 2.01$\pm$0.33 & -0.61$\pm$0.06 \\
$\mathtt{Ms9.3\_C\_\lambda+2.0}$ & 9.3 & 5.8 & 1.98 & -7.04$\pm$0.69 & 20.86$\pm$3.34 & -0.15$\pm$0.24 & 14.47$\pm$2.33 & 3630 & 1.55$\pm$0.25 & 2.24$\pm$0.36 & -0.75$\pm$0.08 \\
$\mathtt{Ms10.6\_C\_\lambda+3.3}$ & 10.6 & 4.1 & 3.27 & -8.51$\pm$1.32 & 22.06$\pm$6.04 & -0.03$\pm$0.24 & 27.41$\pm$4.62 & 5820 & 2.59$\pm$0.44 & 2.08$\pm$0.62 & -0.80$\pm$0.15 \\
$\mathtt{Ms10.1\_C\_\lambda+4.3}$ & 10.1 & 3.7 & 4.33 & -9.32$\pm$1.56 & 24.76$\pm$7.54 & -0.10$\pm$0.31 & 28.28$\pm$7.86 & 23500 & 2.8$\pm$0.78 & 2.45$\pm$0.75 & -0.92$\pm$0.15 \\
\bottomrule
\end{tabular}}
\vspace{0.2in}
\caption{Volume-weighted moments and derived quantities from our compressively-driven simulations. Columns show the run name, volume-weighted and mass-weighted Mach numbers, $\lambda_{\rm a} \equiv \tau_{\rm a}/ \tau_{\rm e}$, the mean, variance, and skewness of $P_{\rm V}(s),$ standard deviation of the density PDF, three fits to the distribution as a function of Mach number ($b_s^2 \equiv [\exp(\sigma_{s,{\rm V}}^2))-1] M_{\rm V}^{-2}$ $b_{\rho}  \equiv \sigma_{\rho} M_{\rm V},$ and $B \equiv \sigma_{\rm V}^2 M_{\rm V}^{-1})$ and two fits to $\left<s\right>_{\rm V}$ as a function of Mach number. We find that its measured value can exceed $10^6$ for many choices of $M_{\rm V}$ and $\tau_{\rm a}.$  Note that we omit uncertainties for $M_{\rm V},$ $M_{\rm M}$ and $\lambda_{\rm a},$ which are always $\approx 10\%,$ as well as for $b_s,$ which is not used in our analysis.}
\label{tab:momentsV_C}
\end{table*}%

\begin{table*}[t]
\hspace{-0.8in}
 \centering
 \resizebox{1.1\textwidth}{!}{
\begin{tabular}{llllcccccccc}
\toprule
& \multicolumn{3}{c}{Flow Properties} & \multicolumn{4}{c}{PDF Moments} & \multicolumn{4}{c}{Derived Values}\\
\cmidrule(lr{.75em}){2-4} \cmidrule(lr{.75em}){5-8} \cmidrule(lr{.75em}){9-12}
Name & $M_{\rm V}$ & $M_{\rm M}$ & $\lambda_{\rm a}$ &
$\left<s\right>_{\rm V}$ & $\sigma_{s,{\rm V}}^2$ & $\mu_{s,{\rm V}}$& $\sigma_{\rho,{\rm V}}$ & $b_s$ & $b_{\rho}$ &
$B$ & $\left<s\right>_{\rm V}$ $M_{\rm V}^{-1}$ \\
$\mathtt{Ms2.0\_M\_\lambda-3.0}$ & 2.0 & 1.9 & -3.01 & -0.59$\pm$0.07 & 1.32$\pm$0.18 & -0.19$\pm$0.19 & 1.31$\pm$0.11 & 0.80 & 0.66$\pm$0.06 & 0.63$\pm$0.09 & -0.30$\pm$0.04 \\
$\mathtt{Ms2.1\_M\_\lambda-1.8}$ & 2.1 & 2.0 & -1.85 & -0.64$\pm$0.07 & 1.47$\pm$0.20 & -0.23$\pm$0.20 & 1.37$\pm$0.11 & 0.84 & 0.65$\pm$0.05 & 0.67$\pm$0.10 & -0.30$\pm$0.03 \\
$\mathtt{Ms2.4\_M\_\lambda-0.5}$ & 2.4 & 2.2 & -0.52 & -0.77$\pm$0.11 & 1.77$\pm$0.37 & -0.32$\pm$0.26 & 1.52$\pm$0.14 & 0.92 & 0.63$\pm$0.06 & 0.74$\pm$0.16 & -0.32$\pm$0.05 \\
$\mathtt{Ms2.5\_M\_\lambda+0.6}$ & 2.5 & 2.1 & 0.60 & -0.75$\pm$0.06 & 1.70$\pm$0.19 & -0.28$\pm$0.15 & 1.51$\pm$0.13 & 0.86 & 0.62$\pm$0.05 & 0.69$\pm$0.08 & -0.30$\pm$0.03 \\
$\mathtt{Ms2.5\_M\_\lambda+1.8}$ & 2.5 & 2.0 & 1.83 & -0.80$\pm$0.05 & 1.80$\pm$0.15 & -0.28$\pm$0.10 & 1.59$\pm$0.10 & 0.91 & 0.64$\pm$0.04 & 0.73$\pm$0.07 & -0.32$\pm$0.02 \\
$\mathtt{Ms2.6\_M\_\lambda+3.0}$ & 2.6 & 2.3 & 2.95 & -0.70$\pm$0.07 & 1.58$\pm$0.20 & -0.27$\pm$0.12 & 1.40$\pm$0.07 & 0.75 & 0.53$\pm$0.03 & 0.60$\pm$0.08 & -0.27$\pm$0.03 \\
[0.3em]
$\mathtt{Ms3.5\_M\_\lambda-0.2}$ & 3.5 & 3.0 & -0.15 & -1.22$\pm$0.11 & 2.96$\pm$0.38 & -0.36$\pm$0.13 & 2.12$\pm$0.16 & 1.22 & 0.61$\pm$0.04 & 0.86 $\pm$0.20 & -0.35$\pm$0.06 \\
[0.3em]
$\mathtt{Ms5.0\_M\_\lambda-2.1}$ & 5.0 & 4.7 & -2.11 & -1.38$\pm$0.15 & 3.07$\pm$0.54 & -0.16$\pm$0.21 & 2.75$\pm$0.21 & 0.92 & 0.56$\pm$0.04 & 0.62$\pm$0.16 & -0.30$\pm$0.05 \\
$\mathtt{Ms4.8\_M\_\lambda-1.0}$ & 4.8 & 4.3 & -1.04 & -1.47$\pm$0.13 & 3.55$\pm$0.58 & -0.33$\pm$0.22 & 2.81$\pm$0.27 & 1.19 & 0.59$\pm$0.06 & 0.74$\pm$0.12 & -0.34$\pm$0.03 \\
$\mathtt{Ms5.0\_M\_\lambda+0.2}$ & 5.0 & 4.1 & 0.21 & -1.71$\pm$0.12 & 4.30$\pm$0.55 & -0.36$\pm$0.23 & 3.10$\pm$0.33 & 1.71 & 0.62$\pm$0.07 & 0.85$\pm$0.12 & -0.34$\pm$0.03 \\
$\mathtt{Ms5.1\_M\_\lambda+1.3}$ & 5.1 & 3.7 & 1.33 & -2.09$\pm$0.12 & 4.96$\pm$0.70 & -0.24$\pm$0.24 & 3.99$\pm$0.31 & 2.33 & 0.78$\pm$0.06 & 0.97$\pm$0.15 & -0.42$\pm$0.03 \\
$\mathtt{Ms5.2\_M\_\lambda+2.6}$ & 5.2 & 4.0 & 2.56 & -2.14$\pm$0.38 & 5.00$\pm$0.92 & -0.17$\pm$0.16 & 3.93$\pm$0.85 & 2.33 & 0.76$\pm$0.16 & 0.96$\pm$0.18 & -0.42$\pm$0.08 \\
$\mathtt{Ms5.4\_M\_\lambda+3.7}$ & 5.4 & 4.0 & 3.66 & -2.33$\pm$0.47 & 5.62$\pm$1.57 & -0.16$\pm$0.21 & 4.24$\pm$1.20 & 3.02 & 0.78$\pm$0.22 & 1.04$\pm$0.31 & -0.45$\pm$0.09 \\
[0.3em]
$\mathtt{Ms7.3\_M\_\lambda+0.6}$ & 7.3 & 5.8 & 0.59 & -2.55$\pm$0.26 & 6.31$\pm$1.57 & -0.17$\pm$0.31 & 4.88$\pm$0.77 & 3.21 & 0.67$\pm$0.11 & 0.86$\pm$0.30 & -0.36$\pm$0.06 \\
[0.3em]
$\mathtt{Ms10.9\_M\_\lambda-1.3}$ & 10.9 & 10.5 & -1.28 & -2.33$\pm$0.15 & 5.57$\pm$0.63 & -0.18$\pm$0.16 & 4.69$\pm$0.46 & 1.48 & 0.43$\pm$0.04 & 0.76$\pm$0.10 & -0.32$\pm$0.03 \\
$\mathtt{Ms10.1\_M\_\lambda-0.3}$ & 10.1 & 9.6 & -0.27 & -2.36$\pm$0.22 & 5.76$\pm$1.01 & -0.25$\pm$0.25 & 4.87$\pm$0.79 & 1.76 & 0.48$\pm$0.08 & 0.53$\pm$0.09 & -0.22$\pm$0.02 \\
$\mathtt{Ms10.9\_M\_\lambda+1.0}$ & 10.9 & 8.6 & 0.99 & -3.01$\pm$0.37 & 8.15$\pm$1.40 & -0.30$\pm$0.13 & 5.44$\pm$0.91 & 5.81 & 0.50$\pm$0.08 & 0.80$\pm$0.14 & -0.30$\pm$0.04 \\
$\mathtt{Ms10.7\_M\_\lambda+2.1}$ & 10.7 & 8.0 & 2.08 & -3.44$\pm$0.36 & 9.60$\pm$1.69 & -0.33$\pm$0.17 & 6.75$\pm$2.06 & 11.2 & 0.63$\pm$0.19 & 0.88$\pm$0.16 & -0.32$\pm$0.04 \\
$\mathtt{Ms11.4\_M\_\lambda+3.3}$ & 11.4 & 7.6 & 3.31 & -3.67$\pm$0.51 & 9.69$\pm$2.07 & -0.25$\pm$0.28 & 8.40$\pm$1.50 & 11.8 & 0.74$\pm$0.13 & 0.90$\pm$0.19 & -0.34$\pm$0.05 \\
$\mathtt{Ms11.2\_M\_\lambda+4.4}$ & 11.2 & 8.2 & 4.42 & -3.24$\pm$0.31 & 9.15$\pm$1.71 & -0.26$\pm$0.23 & 5.37$\pm$0.49 & 8.66 & 0.48$\pm$0.04 & 0.81$\pm$0.16 & -0.29$\pm$0.03 \\
[0.7em]
$\mathtt{Ms3.0\_S\_\lambda-0.3}$ & 3.0 & 2.8 & -0.29 & -0.49$\pm$0.03 & 0.98$\pm$0.07 & 0.04$\pm$0.10 & 1.26$\pm$0.05 & 0.43 & 0.42$\pm$0.02 & 0.32$\pm$0.01 & -0.16$\pm$0.01 \\
[0.3em]
$\mathtt{Ms4.4\_S\_\lambda+0.1}$ & 4.4 & 4.1 & 0.07 & -0.76$\pm$0.05 & 1.57$\pm$0.14 & -0.05$\pm$0.11 & 1.74$\pm$0.07 & 0.44 & 0.40$\pm$0.02 & 0.36$\pm$0.05 & -0.19$\pm$0.02 \\
[0.3em]
$\mathtt{Ms5.3\_S\_\lambda-2.0}$ & 5.3 & 5.0 & -2.02 & -0.87$\pm$0.04 & 1.79$\pm$0.12 & -0.05$\pm$0.12 & 1.96$\pm$0.12 & 0.42 & 0.37$\pm$0.02 & 0.34$\pm$0.03 & -0.17$\pm$0.01 \\
$\mathtt{Ms5.5\_S\_\lambda-0.9}$ & 5.5 & 5.2 & -0.90 & -0.89$\pm$0.05 & 1.81$\pm$0.13 & 0.00$\pm$0.13 & 2.06$\pm$0.09 & 0.41 & 0.38$\pm$0.02 & 0.33$\pm$0.03 & -0.17$\pm$0.01 \\
$\mathtt{Ms6.2\_S\_\lambda+0.4}$ & 6.2 & 5.9 & 0.42 & -0.99$\pm$0.05 & 2.02$\pm$0.13 & -0.02$\pm$0.10 & 2.26$\pm$0.12 & 0.41 & 0.36$\pm$0.02 & 0.32$\pm$0.03 & -0.16$\pm$0.01 \\
$\mathtt{Ms6.5\_S\_\lambda+1.5}$ & 6.5 & 6.2 & 1.55 & -0.99$\pm$0.08 & 2.06$\pm$0.25 & -0.07$\pm$0.11 & 2.20$\pm$0.11 & 0.40 & 0.34$\pm$0.02 & 0.31$\pm$0.04 & -0.16$\pm$0.01 \\
$\mathtt{Ms6.9\_S\_\lambda+2.8}$ & 6.9 & 6.4 & 2.78 & -1.06$\pm$0.08 & 2.22$\pm$0.23 & -0.09$\pm$0.13 & 2.34$\pm$0.12 & 0.42 & 0.34$\pm$0.02 & 0.32$\pm$0.04 & -0.16$\pm$0.01 \\
$\mathtt{Ms7.0\_S\_\lambda+3.9}$ & 7.0 & 6.9 & 3.87 & -1.04$\pm$0.07 & 2.25$\pm$0.27 & -0.22$\pm$0.24 & 2.28$\pm$0.13 & 0.42 & 0.32$\pm$0.02 & 0.32$\pm$0.04 & -0.15$\pm$0.01 \\
[0.3em]
$\mathtt{Ms8.9\_S\_\lambda+0.8}$ & 8.9 & 8.5 & 0.79 & -1.33$\pm$0.10 & 2.99$\pm$0.31 & -0.24$\pm$0.12 & 2.75$\pm$0.14 & 0.49 & 0.31$\pm$0.02 & 0.34$\pm$0.05 & -0.15$\pm$0.02 \\
[0.3em]
$\mathtt{Ms12.3\_S\_\lambda+1.1}$ & 12.3 & 12.2 & 1.09 & -1.55$\pm$0.13 & 3.52$\pm$0.47 & -0.20$\pm$0.12 & 3.09$\pm$0.14 & 0.47 & 0.25$\pm$0.01 & 0.29$\pm$0.06 & -0.13$\pm$0.02 \\
\bottomrule
\end{tabular}}
\vspace{0.2in}
\caption{Volume-weighted moments and derived quantities from our mixed and solenoidally-driven simulations. Columns are as in Table \ref{tab:momentsV_C}}
\label{tab:momentsV_MS}
\end{table*}

\subsection{Volume-Weighted Probability Distributions}
\label{sec:volume}

\subsubsection{Overall Distributions}

Fig.\ \ref{fig:PSV}, shows the volume-weighted probability distribution of $s$ for eight representative groups of simulations. The top row of this figure shows results from fully compressive simulations with Mach numbers of $M_{\rm V} \approx$ 2, 4, 6, and 8 and varying values of $\tau_{\rm a}.$ In all cases, increasing $\tau_{\rm a}$ leads to a systematic broadening of $P_{\rm V}(s)$, and this is associated with the formation of the expanding voids seen in Fig.\ \ref{fig:slices}. This change occurs primarily at low densities, and we see that not only the width of $P_{\rm V}(s)$ increases, but the peak of the distribution is shifted to lower $s$ values.  Note, however, that when $\tau_{\rm a}$ is long, there is also a notable change at high $s$ values, which is due to the presence of shells of swept-up material at the edges of the expanding regions \citep{Paper2}. 

The second row in this figure shows the impact of $\tau_{\rm a}$ on three mixed driving runs with $M_{\rm V} \approx$ 2, 5, and 10, as well as solenoidal runs with $M_{\rm V} \approx 5.$ The mixed cases show hints of same overall trends as the fully compressive runs. However, consistent with the slices shown in Fig.\ \ref{fig:slices}, these changes are much more subtle and often limited to the left tail of the distribution.

Finally, in the solenoidal case, $\tau_{\rm a}$ has no notable impact on $P_{\rm V}(s)$. While the distribution is slightly broader in the large $\tau_{\rm a}$ case, as we shall see below, this change is consistent with the differences in Mach numbers between the runs shown, with the $\tau_{\rm a}=0.01$ run having a Mach number of $M_{\rm V}=5.3$ and the $\tau_{\rm a}=3$ run having a Mach number of $M_{\rm V}=7.5.$  The mean, variance, and skewness of these distributions are quantified in Table \ref{tab:momentsV_C} for the compressive runs and in Table \ref{tab:momentsV_MS} for the mixed and solenoidal runs.

\begin{figure*}[t]
\begin{center}
\includegraphics[width=\textwidth]{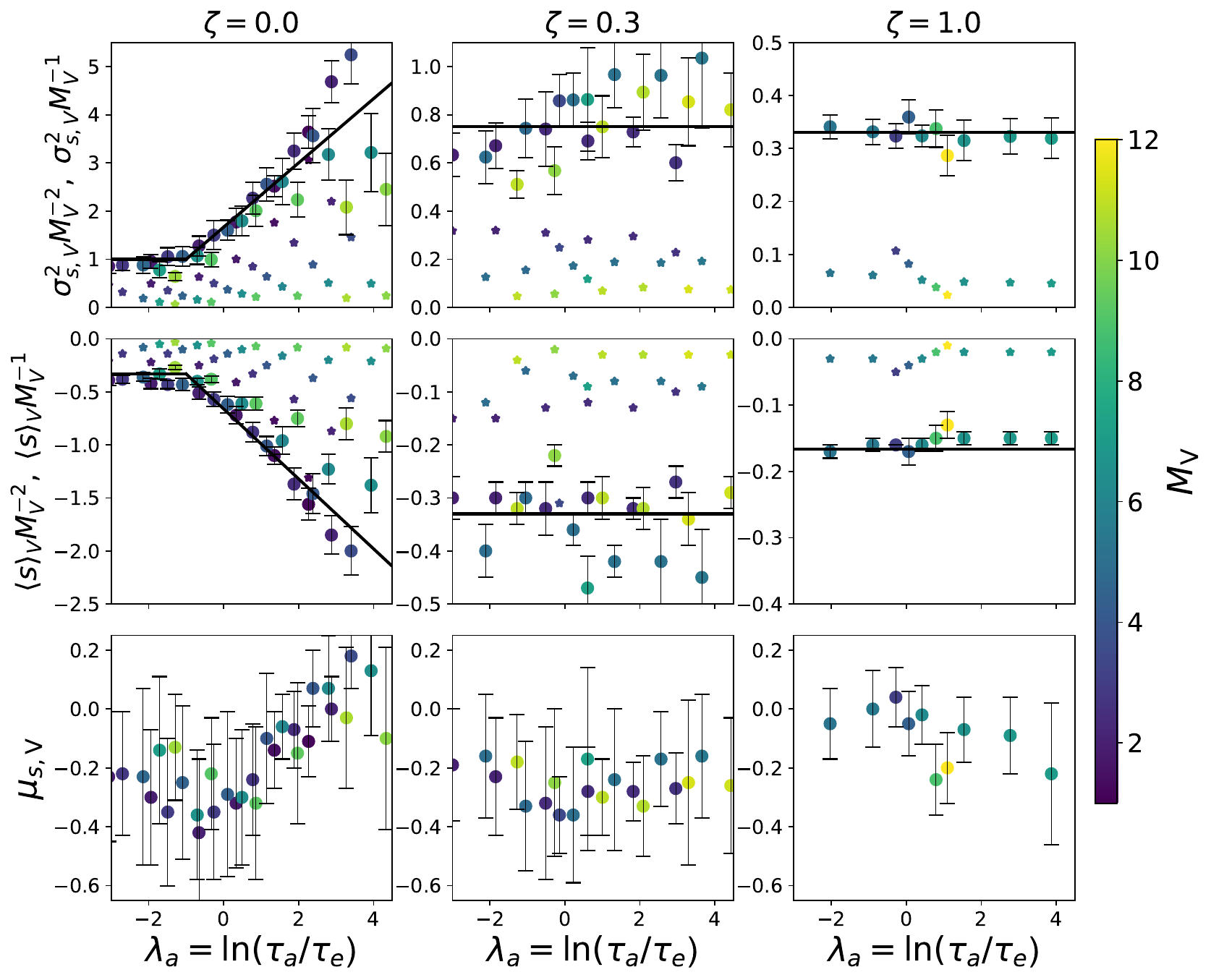}
\end{center}
\vspace{-0.1in}
\caption{{\em Top:} The volume-weighted variance of $s$, normalized by the volume-weighted Mach number squared, $\sigma^2_{s,{\rm V}} M_{\rm V}^{-2},$ (stars) and by the volume-weighted Mach number, $B \equiv \sigma^2_{s,{\rm V}} M_{\rm V}^{-1},$ (circles), which provides a much better description of the data. In all cases, the colors correspond to the Mach number. Columns show results from compressively-driven turbulence (left), mixed-driving turbulence (center), and solenoidally-driven turbulence (right).  The solid lines are fits $B = 1 + \frac{2}{3}(1+ \lambda_{\rm a}) \Theta(1+\lambda_{\rm a})$, $B= \frac{3}{4},$ and $B=\frac{1}{3},$  where $\lambda_{\rm a} \equiv \ln(\tau_{\rm e}/\tau_{\rm a})$ and $\Theta$ is the Heaviside step function. {\em Middle:} The volume-weighted mean values of $s$, normalized by the volume-weighted Mach number squared, $\left< s \right>_{\rm V} M_{\rm V}^{-2},$ (stars) and by the volume-weighted Mach number, $\left< s \right>_{\rm V} M_{\rm V}^{-1},$ (circles). Again, normalizing by $M_{\rm V}^{-1}$ provides a better description of the data, and from left to right the lines give fits of $\left<s\right>_{\rm V} M_{\rm V}^{-1}=-\frac{1}{3}- \frac{1}{3}( 1+\lambda_{\rm a}) \Theta(1+\lambda_{\rm a}),$ $\left<s\right>_{\rm V} M_{\rm V}^{-1}=-\frac{1}{3}$, and $\left<s\right>_{\rm V} M_{\rm V}^{-1}=-\frac{1}{6}.$ {\em Bottom:} Skewness of $P_{\rm V}(s).$ For most cases, the distributions are negatively skewed, although $\mu_{s,\rm V}$ is the largest in the compressively-driven and mixed simulations with small driving correlation times. Unlike $\sigma_{s,{\rm V}}$ and $\left<s\right>_{\rm V}$, skewness shows no strong trends with Mach number.}
\label{fig:fitV}
\end{figure*}

\subsubsection{Impact of Mach Number on $\sigma_s$ and $\sigma_\rho$}

In Figure \ref{fig:sigrho_sig2s}, we show the variance of $\rho$ and $s$ for these distributions. In the top row of this Figure, we plot the behavior of $\sigma^2_{s,{\rm V}}$ as a function of $M_{\rm V}$ in our simulations, as well as results from previous studies.  As discussed above, the most widely applied fit relating the variance of $s$ and the Mach number is given by eq.\ (\ref{eq:standardfit}),  $\sigma_{s,{\rm V}}^2 = \ln(1+b_s^2 M_{\rm V}^2).$ In the literature, $b_s$ is often assumed to be $\approx \frac{1}{3}$ for solenoidally-driven turbulence and $\approx 1$ for compressively-driven turbulence \citep[e.g.][]{Ostriker2001, MacLow05, Kowal07, Lemaster08,Federrath08,Price11,Burkhart12,Seon12,Lee20,Hennebelle24}.

These fits are shown as the dashed lines in Fig.\ \ref{fig:sigrho_sig2s}, with $b_s$ values as quoted in the panels.  As expected from the volume-weighted PDFs in Fig.\ \ref{fig:PSV}, there is a clear systematic increase of $\sigma_{s,{\rm V}}^2$ with $\lambda_{\rm a}$ in the fully compressive runs, such that $\sigma_{s,{\rm V}}^2$ cannot be well fit as a pure function of $M_{\rm V}.$  Furthermore, the standard expression with $b_s=1$ falls below all the measured $\sigma_{s,{\rm V}}^2$ values.  While increasing this to $b_s = 2$ allows us to capture the variance of the runs with driving correlation times $\tau_a \lesssim 0.1 \tau_e,$ even this value strongly underestimates the variance of runs with larger driving correlation times. Also in the mixed-driving runs with $\zeta=0.3$, the standard fit fails to reproduce the data, even for the runs with the smallest driving correlation times.

Tables \ref{tab:momentsV_C} and \ref{tab:momentsV_MS} present several derived values that connect $\sigma_{s,{\rm V}}^2$ and $\left<s\right>_{\rm V}$ with $M_{\rm V}.$ Inverting eq.\ (\ref{eq:standardfit}) gives
\begin{equation}
b_s \equiv \left[{\rm exp} \left(\sigma_{s,{\rm V}}^2 \right) - 1 \right]^{1/2} M_{\rm V}^{-1} .
\label{eq:bs}
\end{equation}
From these tables, we see that, for large values of $\tau_{\rm a}$ and $M_{\rm V},$ $b_s$ can exceed 10 in the mixed case, and reach values exceeding $1000$ in some of the compressive runs. Thus, we conclude that when compressive driving is significant, eq.\ (\ref{eq:standardfit}) does not provide a good description of the variance of $s$ as a function of $M_{\rm V}$ and $\lambda_{\rm a}.$

The top right panel of Fig.\ \ref{fig:sigrho_sig2s} shows $\sigma_{s,{\rm V}}^2$ from solenoidal runs, which, as expected, does not depend on the driving correlation time. In this case, which is the one most studied in previous simulations, eq.\ (\ref{eq:standardfit}) provides a good fit to the data, with $b_s$ somewhere in the range of $\frac{1}{3}$ to $\frac{2}{5}.$ 

There are two assumptions underlying eq.\ (\ref{eq:standardfit}). The first is that $\sigma^2_{\rho,V} \propto M^2_{\rm V},$ which is motivated by the fact that the density contrast behind an isothermal shock is proportional to $M_{\rm V}^2$, but the shocked gas occupies only a fraction $M_{\rm V}^{-2}$ of the original volume \citep{Padoan97}. The second assumption is that $P_{\rm V}(s)$ is Gaussian, which is violated to varying extents across runs, as quantified by the significant negative skewness values shown in Tables \ref{tab:momentsV_C} and \ref{tab:momentsV_MS}.

To test the assumption that $\sigma_{\rho,V} = b_\rho M_{\rm V},$ we plot $\sigma_{\rho,V}$ as a function of $M_{\rm V}$ in the second row of Fig.\ \ref{fig:sigrho_sig2s}.  Note that this is the standard deviation and not the variance, and we use the subscript $\rho$ to denote that this is a fit to the distribution of the density rather than the log-density.   Here we consider values of $b_\rho$ equal to the values of $b_s$ in the upper panels, and we see that, in general, $\sigma_{\rho,V} = b_\rho M_{\rm V}$ provides a reasonable fit to our simulation results.  

In the fully compressive case, $b_\rho=1$ provides a good description of runs with $\tau_{\rm a} \approx \tau_{\rm e},$ while runs with longer $\tau_{\rm a}$ values are more closely fit with $b_\rho=2.$ Similarly, in the $\zeta=0.3$ case, $\sigma^2_{\rho,V} = b_\rho^2 M^2_{\rm V}$ provides an approximate description of the data, with $b_\rho$ between $\frac{1}{2}$ and $\frac{3}{4}$, depending on the driving correlation time.  Interestingly, the one case that is clearly discrepant with $\sigma^2_{\rho,V} \propto M^2_{\rm V}$ is the solenoidal one, which drops below the $b_\rho = \frac{1}{3}$ line at high Mach numbers \citep{Price11}. The origin of the drop is perhaps due to the fact that shocks are underresolved in simulations for very large Mach numbers.     

Our $\sigma_{\rho,V}$ measurements show that the primary issue with eq.\ (\ref{eq:standardfit}) is that $P_{\rm V}(s)$ cannot be adequately approximated by a Gaussian. To quantify this, we plot $\sigma^2_s$  as a function of $\sigma_{\rho,V}$ in the bottom row of Fig.\ \ref{fig:sigrho_sig2s}. If $P_{\rm V}(s)$ were Gaussian, these quantities would be related as $\sigma^2_{s,{\rho,V}}= \ln(1+\sigma_{\rho,{\rm V}}^2),$ which is given by the dashed lines in the panels.  Here we see that our simulation results, along with others in the literature, depart strongly from the expectation in all cases in which $\sigma_{\rho,V} \gtrsim 2.$  Instead, the relation between $\sigma^2_{s,{\rm V}}$  and $\sigma_{\rho,V}$ is much closer to linear, as given by solid lines, although $\sigma^2_{s,{\rm V}}$ exceeds even this relation in compressive cases in which $\lambda_{\rm a}$ is large.

\subsubsection{Impact of Mach Number and $\tau_a$ on $P_{\rm V}(s)$}

\begin{figure*}[t]
\begin{center}
\includegraphics[width=\textwidth]{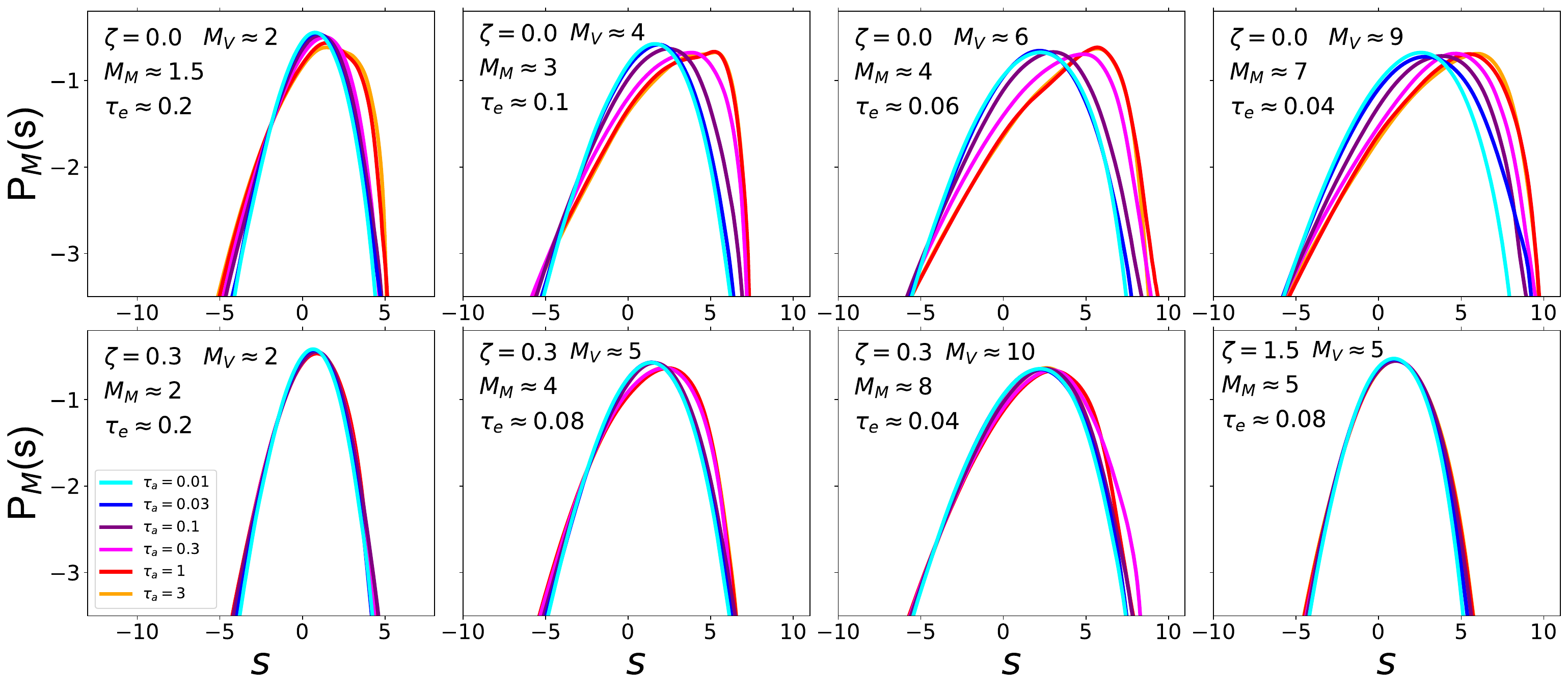}
\end{center}
\vspace{-0.2in}
\caption{Mass-weighted PDFs from a representative subset of our simulations. Columns, rows, and line styles are as in Fig.\ \ref{fig:PSV}. For the compressive runs, increasing $\tau_{\rm a}$ shifts the peak to the right, broadens the distribution and leads to a large negative skewness. As in the volume-weighted case, similar effects are seen to a limited degree in the mixed driving runs, while the differences between the solenoidal runs are consistent with small changes in the Mach number.}
\label{fig:PSM}
\end{figure*}

Fig.\ \ref{fig:fitV} shows the dependence of the moments of $P_{\rm V}(s)$ on the Mach number and driving correlation time.  Given the strong discrepancy between  eq.~(\ref{eq:standardfit}) and our results, the top panel of this figure shows $\sigma^2_{s,{\rm V}}$ normalized instead by powers of the Mach number. Here we see $\sigma^2_{s,{\rm V}}  \propto M_{\rm V}^2,$ is too strong a dependence on Mach number to be a good fit to the data, and that the $\sigma^2_{s,{\rm V}}$ dependence on $\lambda$ is a not linear.   Adopting a functional fit of the form $\sigma^2_{s,{\rm V}}  = M_{\rm V}^\alpha  [1 + \beta (1+\lambda_a) \theta(1+\lambda_a)],$ and carrying out a least squares fit to our measurements, we find a best fit value of $\alpha=0.85.$ Since this is very close to $\alpha=1,$ choosing a scaling with Mach number more complicated than $\sigma^2_{s,{\rm V}}$  proportional to $M_{\rm V}$ is not justified.

Thus, we define a fit parameter 
\be
 B \equiv \sigma^2_{s,{\rm V}} M_{\rm V}^{-1},
\ee
which is given as the third column of derived values in the table. As we can see from Fig.\ \ref{fig:fitV}, this greatly reduces the difference between the runs, such that $B$ is simply a function of the driving correlation time.   

Having fixed the dependence on Mach number, we find that the best fit $\beta$ value is $0.65.$
Thus the dependence of the variance on $\tau_{\rm a}$ can be well reproduced by a piecewise linear function: 
\be
 B \approx 1 + \frac{2( 1+\lambda_{\rm a})}{3} \Theta(1+\lambda_{\rm a}),
\ee
where $\Theta$ is the Heaviside step function. 
This means that for fully-compressive driving, $\sigma^2_{s,{\rm V}}$ is equal to $M_{\rm V}$ for small values of the driving correlation time, and it grows linearly with $\lambda_{\rm a}$ whenever $\lambda_{\rm a} > -1.$  

Physically, this can be understood to occur because for very small values of $\tau_{\rm a}$ the driving pattern changes many times per eddy turnover time, so that the production of large voids is minimal, and $P_{\rm V}(s)$ is independent of the driving correlation time. On the other hand, as $\tau_{\rm a}$ grows and the expansions that give rise to voids are sustained longer, $P_{\rm V}(s)$ broadens significantly, with an overall dependence that is proportional to $\lambda_{\rm a} \equiv \ln(\tau_{\rm a}/\tau_{\rm e}).$  In the upper left panel of Fig.\ \ref{fig:sigrho_sig2s}, we plot $\sigma^2_{s,{\rm V}} = B M_{\rm V}$ for a value of $B=2.$

The linear growth of $B$ with $\lambda_{\rm a}$ suggests that as $\tau_{\rm a}$ approaches infinity, $\sigma^2_{s,{\rm V}}$  diverges logarithmically. To test this, we carried out an additional $M_{\rm V} \approx 3.5$ simulation with static, purely compressive driving, and found that $\sigma^2_{s,{\rm V}}$ kept increasing logarithmically to the final output at $\approx 100 \tau_e$. \cite{Pan22} showed that in simulations with solenoidal driving, the density power spectrum is determined by the balance between the pseudosound and acoustic effects and the pressure term. For compressive driving, the driving acceleration may directly affect the density spectrum and cause an imbalance such that a steady state may not be achieved for a static driving pattern. This will be investigated in a separate work.

The center left panel of Fig.\ \ref{fig:fitV} presents $\left<s\right>_{\rm V} M_{\rm V}^{-2}$ and $\left<s\right>_{\rm V} M_{\rm V}^{-1}$ as a function of $\lambda_{\rm a}$ from the compressive runs. For an exactly lognormal PDF, mass conservation requires that $\left<s\right>_{\rm V} = -\sigma_{s,{\rm V}}^2/2.$ However, as the actual distribution is skewed slightly to low $s$ values, this shifts the mean value to larger values.  Carrying out a similar least squares fit as we did for $\sigma_{s,{\rm V}}^2,$ we find that
\be
\left<s\right>_{\rm V}M_{\rm V}^{-1}\approx -\frac{B}{2} + \frac{1}{6} = -\frac{1}{3} -\frac{ 1+\lambda_{\rm a}}{3} \Theta(1+\lambda_{\rm a})
\ee
provides a good description of this data.

The skewness of $P_{\rm V}(s)$ from the compressive runs is shown in the lower left panel of Fig.\ \ref{fig:fitV}. This quantifies the asymmetrical tails of the distribution, which are biased to low $s$ values as seen in Fig.\ \ref{fig:PSV}. There is a mild trend of $\mu_{s,\rm V}$ becoming more negative between $\lambda_{\rm a} \approx -3$ and $\approx 0$ and then approaching $\approx 0$ at large $\lambda_{\rm a}$ values but the overall scatter is large and so we do not attempt to provide a fit to this trend.

The center column of Fig.\ \ref{fig:fitV} presents the results from our mixed-driving simulations, which correspond to the numbers given in Table \ref{tab:momentsV_MS}. As in the compressive case, the Mach number dependence in these simulations is much better fit by $\sigma^2_{s,{\rm V}}$ and $\left<s\right>_{\rm V}$ $\propto M_{\rm V}$ than to $M_{\rm V}^2.$ However, In the mixed-driving case, the overall variance of $s$ is smaller and the dependence of $\sigma^2_{s,{\rm V}}$ on $\lambda_{\rm a}$ is much weaker, because the compressions and expansions caused directly by the driving are much smaller.  For $\zeta = 0.3,$ we find that a constant value of $\sigma^2_{s,{\rm V}} \approx \frac{3}{4} M_{\rm V}$ provides an adequate match to the data, and that more complicated fits are not warranted given the uncertainties in our measurements. Likewise, a value of $\left<s\right>_{\rm V}M_{\rm V}^{-1} \approx - \frac{1}{3}$ provides a good fit to the average value of  $s$ and the skewness is $\mu_{s,\rm V} \approx 0.3$ in all cases.

Finally, the right column of Fig.\ \ref{fig:fitV} shows the results of our solenoidal simulations.  As discussed in more detail in \cite{Paper2}, the driving correlation time is not likely to have an indirect impact on the density distribution in these simulations. This is because the density distribution is set by small-scale changes, and is thus insensitive to the pattern of large-scale driving.  In this case, $B\approx \frac{1}{3}$ provides a good fit to all simulations, which we also plot in the upper central panel of Fig.\ \ref{fig:sigrho_sig2s}.  This expression is also very close to eq.\ (\ref{eq:standardfit}) when $M_{\rm V} \lesssim 8.$  In the solenoidal case, the mean and skewness are well described by $\left<s\right>_{\rm V} M_{\rm V}^{-1}\approx -\frac{1}{6},$ and $\mu_{s,\rm V} \approx 0.1$ respectively.

\subsection{Mass-Weighted Probability Distributions}
\label{sec:mass}

\subsubsection{Overall Distributions}

Next, we consider the mass-weighted probability distribution of $s,$ which is shown in Fig.\ \ref{fig:PSM} for the same set of eight representative groups of simulations shown in Fig.\ \ref{fig:PSV}. Although the mass-weighted PDF can be calculated directly from the volume-weighted PDF as $P_{\rm M}(s) = \frac{\rho}{\rho_0} P_{\rm V}(s) = e^{s} P_{\rm V}(s),$ it emphasizes different features, providing a complementary viewpoint.

\begin{figure*}[t]
\begin{center}
\includegraphics[width=\textwidth]{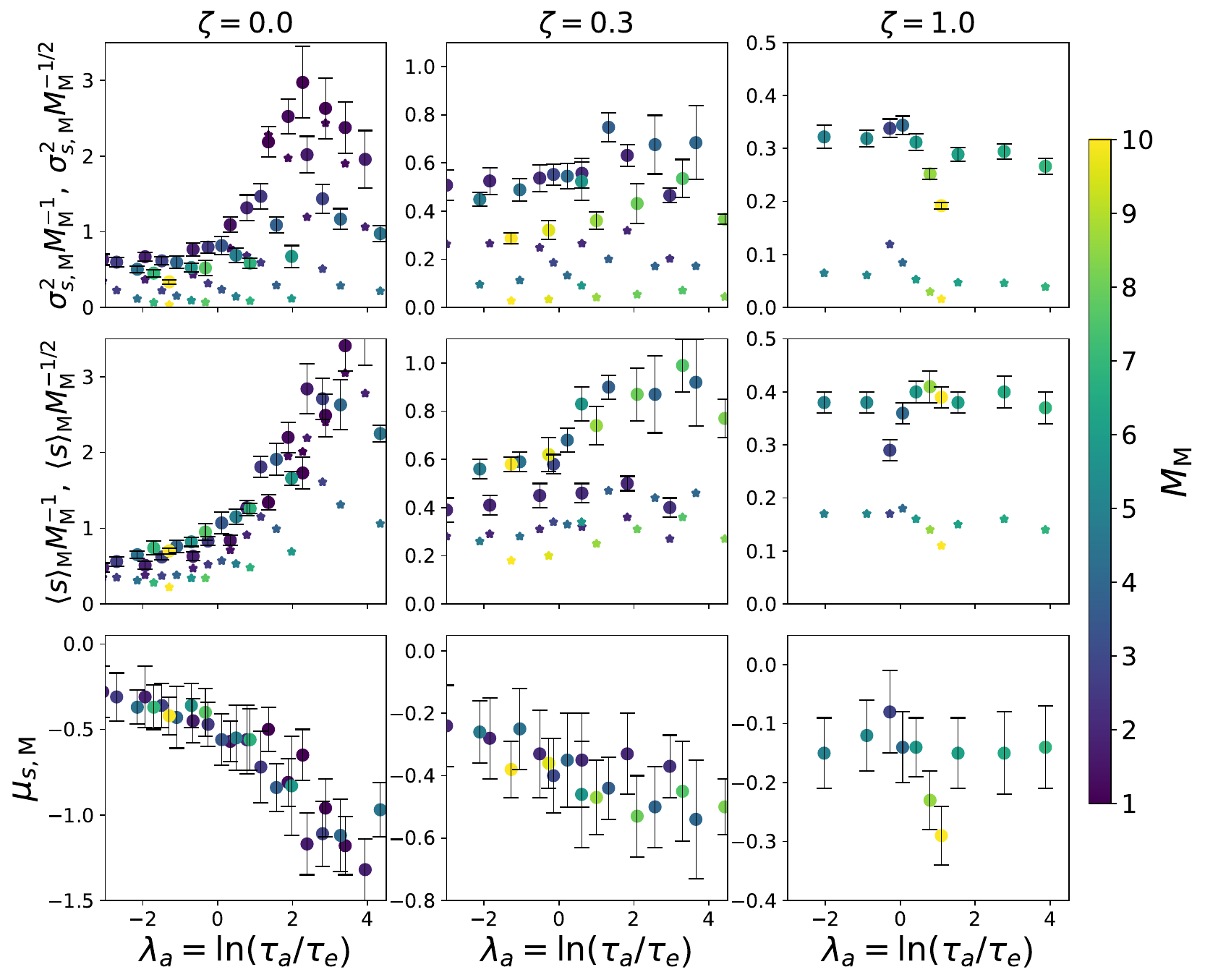}
\end{center}
\vspace{-0.1in}
\caption{{\em Top:} The mass-weighted variance of $s$, normalized by the mass-weighted Mach number, $ \sigma^2_{s,{\rm M}} M_{\rm M}^{-1,}$ (stars) and by the square root of the mass-weighted Mach number, $\sigma^2_{s,{\rm M}} M_{\rm M}^{-1/2}$ (circles). Columns show results from compressively-driven turbulence (left), mixed-driving turbulence (center), and solenoidally-driven turbulence (right). While normalizing by $M_{\rm M}^{-1/2}$ reduces the scatter more than normalizing by $M_{\rm M}^{-1},$ the relation is always noisy, so we do not attempt to fit it. {\em Middle:} The mass-weighted variance mean values of $s$, normalized by the mass-weighted Mach number $\left< s \right>_{\rm M} M_{\rm M}^{-1}$ (circles) and by the square root of the mass-weighted Mach number $\left< s \right>_{\rm M} M_{\rm M}^{-1/2}$ (stars). Again, normalizing by $M_{\rm V}^{-1/2}$ reduces the scatter, but both relations are noisy. {\em Bottom:} Skewness of $P_{\rm M}(s).$ All the distributions are negatively skewed, and $\mu_{\rm M}$ is the largest in the compressively-driven simulations with long driving correlation times. The mass-weighted skewness shows no strong trends with Mach number, but in the compressive and mixed cases $\mu_{\rm M}$ becomes much more strongly negative as $\lambda_{\rm a}$ increases.}
\label{fig:fitMM}
\end{figure*}

As in Fig.\ \ref{fig:PSV}, the top row of Fig.\ \ref{fig:PSM} shows results from fully compressive simulations with increasing Mach numbers, while the bottom row shows three groups of mixed-driving cases and a single group of solenoidal runs. In the compressive cases, increasing $\tau_{\rm a}$ leads to a systematic broadening of $P_{\rm M}(s)$ similar to that seen for $P_{\rm V}(s),$ but in the mass-weighted case, this broadening shifts the peak to higher $s$ values. For a lognormal distribution, mass conservation requires this shift to go as $\left<s\right>_{\rm M} = \sigma_{s,{\rm M}}^2/2,$ but, as is clear in Table 5, this is pushed to $\left<s\right>_{\rm M} > \sigma_{s,{\rm M}}^2/2,$ by the overall negative skewness of the distribution.

At the same time, $P_{\rm M}(s)$ is subject to a strong downturn at high $s$ values. As discussed in \cite{Paper1}, this is due to the high thermal pressure at these densities, which exceeds the ram pressure of material at more typical $s$ values moving at $M_{\rm V} c_s.$ This causes shocks to decelerate to subsonic speeds as they move into these regions, drastically reducing their ability to compress the material to even higher densities. As a result, at large $\tau_{\rm a}$ values, $P_{\rm M}(s)$ becomes highly skewed; overdense parcels of gas are found over a tight range of $s$ values that narrows as $\tau_{\rm a}$ increases.  At the same time, due to the presence of expanding voids, underdense parcels of gas are found over a wide range of $s$ values that broadens with increasing $\tau_{\rm a}.$ 

\subsubsection{Impact of Mach Number and $\tau_a$ on $P_{\rm M}(s)$}

These trends are plotted in the left column of Fig.\ \ref{fig:fitMM}, with corresponding values given in Table \ref{tab:momentsM_C} in the appendix. Here, the points with errorbars in the upper left panel show $\sigma^2_{s,{\rm M}} M_{\rm M}^{-1},$ which is analogous to the quantity $B \equiv \sigma^2_{s,{\rm M}} M_{\rm V}^{-1},$ discussed above. Note that as $\sigma^2_{s,{\rm M}}$ is a mass-weighted quantity, we choose to normalize it by the mass-weighted Mach number, although, as shown in the Appendix, this choice has no impact on our overall conclusions.

Like the volume-weighted case, Fig,\ \ref{fig:fitMM} shows a strong trend of $\sigma_{s,{\rm M}}$ increasing at longer driving correlation times, indicating that the voids present in these simulations have a strong impact on the mass distributions. Unlike the volume-weighted case, however, normalizing by $M_{\rm M}$ does not provide a good description of our results, and we find that the large $M_{\rm M}$ cases having the lowest $\sigma^2_{s,{\rm M}} M_{\rm M}^{-1}.$ To attempt to correct for this, we also show values of $\sigma^2_{s,{\rm M}} M_{\rm M}^{-1/2}.$ While this normalization reduces the scatter between simulation results somewhat better than scaling by $M_{\rm M},$ the relation is much noisier than in the volume-weighted case, and for this reason, we do not attempt to fit it with a simple function.
 
The center left panel of Fig.\ \ref{fig:fitMM} shows the mass-weighted average $s$ from our compressive simulations. As is true for the volume-weighed average values, $\left<s\right>_{\rm M}$ is determined by mass conservation when $P_{\rm M}(s)$ is exactly lognormal, such that $\left<s\right>_{\rm M} = \sigma_{s,{\rm M}}^2/2$ in this case. In fact, the mass-weighted mean and variance  are likely to scale similarly with Mach number, even when $P_{\rm M}(s)$ is significantly skewed, as is true for many of the cases shown in Fig.\ \ref{fig:PSM} . Thus, to match $\sigma_{s,{\rm M}}^2,$ we normalize $\left<s\right>_{\rm M}$ by $M_{\rm M}$ and $M_{\rm M}^{1/2}$ in this panel.

As was the case with the variance, normalizing $\left<s\right>_{\rm M}$ by $M_{\rm M}^{1/2}$ reduces the scatter between simulation results somewhat better than scaling by $M_{\rm M}.$ Also like the variance, the scatter remains large enough that we do not attempt to fit these results with a simple function. Nevertheless, it is still clear that increasing $\tau_{\rm a}$ has a strong impact on $\left<s\right>_{\rm M}.$

The lower left panel of this figure shows the mass-weighted skewness in our compressive simulations. Unlike in the volume-weighted case, the correlation between $\mu_{s,{\rm M}}$ and $\lambda_{\rm a}$ is extremely strong, such that in cases with the largest $\lambda_{\rm a}$ values, $\mu_{s,{\rm M}}$ is below $-1$, meaning that $\left< \left( s - \left<s\right>_{\rm M} \right)^3\right> < - \sigma^3_{s, M}(s)$. In these cases, the material with densities above $\left<s\right>_{\rm M}$ is found at a relatively small range of $s$ values, which is capped at $\approx \left<s\right>_{\rm M} + 3 \sigma_{s,{\rm M}}$. 

In \citet{Paper1}, we showed that this occurs because shocks decelerate as they move into denser regions, such that the typical Mach number of a shock approaches $1$ in the densest regions. Calculating the average Mach number as a function of $s$ across our simulation suite, we again find a strong trend of $M$ with $s$, and although there is a significant scatter between runs, typically $M \approx 1$ when $s$ is $\approx \left<s\right>_{\rm M} + 3 \sigma_{s,{\rm M}}$, corresponding to the downturn in $P(s)$.

In fact, this downturn is likely the cause of $\mu_{s,{\rm M}}$ becoming more negative with increasing $\lambda_{\rm a}.$ As the driving correlation time goes up, the distribution gets broader and $\left<s\right>_{\rm M}$ shifts to higher values, but the downturn stays largely fixed.  At the same time, this shift of  $\left<s\right>_{\rm M}$ results in a longer tail of the distribution to low $s$ values.  This leads to an increasingly asymmetric distribution, characterized by more negative values of the skewness.

Similar trends are seen in the mixed ($\zeta=0.3$) simulation results, shown in the central column of Fig.\ \ref{fig:fitMM}. In this case, $\sigma^2_{s,{\rm M}} M_{\rm M}^{-1}$ and $\sigma^2_{s,{\rm M}} M_{\rm M}^{-1/2},$ show a similar scatter, and there is no reason to prefer one normalization over the other. In both cases, however, a significant, but mild increase of $\sigma^2_{s,{\rm M}}$ with $\lambda_{\rm a}$ is seen, indicating that the impact of sustained compressions and expansions is still present, but much less effective than in the $\zeta=0$ case. 

When $\zeta=0.3,$ $\left<s\right>_{\rm M}$ also shows a gradual increase at high $\lambda_{\rm a}$ values, as expected due to the fact that this quantity is strongly correlated with $\sigma_{s,{\rm M}}$ due to mass conservation. Like the purely compressive case, the strongest trend  seen when $\zeta =0.3$ is the correlation between skewness and $\lambda_{\rm a},$ which moves from $\mu_{s,{\rm M}} \approx -0.2$ when $\lambda_{\rm a} \approx -3$ to $\mu_{s,{\rm M}} \approx -0.6$ when $\lambda_{\rm a} \approx 4.$ 

Finally, in the solenoidally-driven ($\zeta=1.0$) case, there is a similar scatter when $\sigma^2_{s,{\rm M}}$ and $\left< s \right>_{\rm M}$ are normalized by $M_{\rm M}$ as when they are normalized by $M_{\rm M}^{1/2}.$ As expected, when turbulence is driven purely solenoidally, the driving correlation time has no measurable effect on the mean, variance, and skewness of the distribution. As $\nabla \cdot {\bf a} = 0$, the force does not directly influence $\nabla \cdot {\mathbf v}$ or the change in density. Instead, compressions and expansions occur indirectly, due to nonlinear interactions, which act stochastically and whose impact is purely dependent on the overall Mach number.

\section{Conclusions}

Supersonic turbulence plays a key role in determining the structure and evolution of astrophysical systems over a wide range of scales. It is driven by both solenoidal and compressive forces, producing a density distribution that depends on the strength of the forcing and the mix of these two types of modes. In cases in which efficient cooling leads to a nearly isothermal equation of state, this density distribution is often approximated as lognormal, with a variance that goes as $\sigma_{s,{\rm V}}^2 \approx \ln \left(1 + b_s^2 M_{\rm V}^2 \right)$, where $s \equiv \ln \rho/\rho_0$, $M_{\rm V}$ is the rms volume-weighted Mach number, and $b_s$ is a constant that depends on the mix of solenoidal and compressive modes.

However, this approximation has several key limitations. First, it neglects the significant skewness of $P_{\rm V}(s)$, particularly in the case in which turbulence is driven compressively. Secondly, it assumes a relation between $\sigma_{s,{\rm V}}^2$ and $M_{\rm V}$ that fails in cases in which the Mach number is large and/or the compressive driving is significant. Finally, and most importantly, it ignores the fact that $M_{\rm V}$ and the compressive fraction are not the only parameters that set the density distribution.

As demonstrated in \cite{Paper2}, when compressive driving is significant, the correlation time of driving accelerations, $\tau_{\rm a},$ also plays a critical role. This is because if $\tau_{\rm a}$ is comparable to the eddy turnover time, $\tau_{\rm e}$, sustained expansions produce large, low-density voids, while these voids are suppressed when  $\tau_{\rm a} \ll \tau_{\rm e}$. Together, these findings show that the density structure of supersonic turbulence cannot be captured by eq.\ (\ref{eq:standardfit}).

In this work, we address this issue by conducting a suite of simulations spanning a wide parameter space of Mach numbers, driving mechanisms, and driving correlation times. Our key findings are as follows:
\begin{itemize}

 \item Over a wide range of parameter space, the relation between the variance of $s$ and the standard deviation of $\rho$ differs strongly from $\sigma_{s,{\rm V}}^2 = \ln (1+\sigma_{\rho,V}^2),$ as occurs in a Gaussian distribution. Instead the relation between $\sigma_{s,{\rm V}}^2$  and $\sigma_{\rho,V}$ is much closer to linear.
 
 \item Compressively-driven turbulence exhibits a strong dependence on $\tau_{\rm a},$  such the dependence of the variance of $s$ on Mach number and $\tau_{\rm a}$ is well described by
 $$\sigma_{s,{\rm V}}^2 \approx M_{\rm V} \left[1 + \frac{2}{3}(1 + \lambda_{\rm a})\Theta(1 + \lambda_{\rm a})\right],$$
 where $\lambda_{\rm a} \equiv \ln(\tau_{\rm a}/\tau_{\rm e})$ and $\Theta$ is the Heaviside step function. For $\tau_{\rm a} \ll \tau_{\rm e}$, the variance simplifies to $\sigma_{s,{\rm V}}^2 \approx M_{\rm V}$, while for $\tau_{\rm a} \gtrsim \tau_{\rm e}$, it grows linearly with $\lambda_{\rm a}$.
 
\item Mixed-driven turbulence shows a much weaker dependency on $\tau_a$ such that,  when $\zeta = 0.3$  our measurements are consistent with
$$\sigma_{s,{\rm V}}^2 \approx  \frac{2}{3} M_{\rm V}.$$
 
\item
 In solenoidally-driven turbulence, $\sigma$ is independent of the driving correlation time because, in this case, there is no direct impact of the driving motions on changes in density. The results of the fully solenoidal-driving are will fit by 
$$\sigma_{s,{\rm V}}^2 \approx \frac{1}{3}M_{\rm V}.$$
Unlike the compressive mixed cases, when $M_{\rm V}\lesssim 8$ this is very similar to the standard relation of $\sigma^2_{s,{\rm V}}=\ln(1+b_s^2 M_{\rm V}^2)$ with $b_s = 1.$
  
 \item The volume-weighted mean, $\langle s \rangle_{\rm V}$, exhibits systematic trends consistent with the variance, as required by mass conservation. In all cases, the volume-weighted skewness is small but significant, with $\mu_{s,{\rm V}} \approx -0.2$ for most of the cases we simulated.

 \item Like the volume-weighted PDF, the mass-weighted PDF, $P_{\rm M}(s),$ becomes much broader for compressively-driven turbulence when $\tau_{\rm a} \gtrsim \tau_{\rm e}.$ Also like the volume-weighted PDF, $P_{\rm M}(s),$  becomes weakly broader when $\zeta=0.3$ and it is independent of $\tau_{\rm a}$ for solenoidally-driven turbulence. However, unlike the volume-weighted results, $\sigma_{s,{\rm M}}$ and $\left< s \right>_{\rm M}$ exhibit a Mach number dependence that is somewhat shallower than $\propto M$ and their dependence on the driving correlation time cannot be reduced to simple piecewise functions.
 
 \item The strongest trend seen in $P_{\rm M}(s)$ is an increase in the skewness of the distribution in compressively-driven cases, particularly those with long driving correlation times. In these cases, the material with densities above $\left<s\right>_{\rm M}$ is found in a relatively small range of $s$ values, which is capped by the value at which the thermal pressure of the gas is comparable to the ram pressure of a typical shock. At the same time, the material with $s$ below $\left<s\right>_{\rm M}$ is found over a large range of values, as the presence of large voids leads to regions in which the density is orders of magnitude below the mean.
\end{itemize}

Together, our results provide an improved description of the behavior of stochastically-driven, supersonic isothermal turbulence and they provide a refined framework for studying turbulence in astrophysical systems. These results are particularly important for systems in which compressive driving plays and a strong role, such as the interstellar medium and star-forming molecular clouds. Note, however, that in such systems, turbulence may be driven by multiple processes with distinct correlation times. For example, as estimated in \cite{Paper2}, stellar feedback may be associated with relatively short $\tau_{\rm a}$ values, leading to a narrow PDF, while gravitational collapse could produce longer $\tau_{\rm a}$ values, corresponding to broader distributions. Future work should explore these effects, allowing for more accurate analyses of highly-turbulent astrophysical systems.
 
\section*{Acknowledgments}

ES acknowledges support from NASA grants 80NSSC22K1265, 80NSSC23K0646, ,and 80NSSC25K7299. PG acknowledges funding by the Deutsche Forschungsgemeinschaft (DFG, German Research Foundation) – 555983577. MB acknowledges support from the Deutsche Forschungsgemeinschaft under Germany's Excellence Strategy - EXC 2121 "Quantum Universe" - 390833306 and from the BMBF ErUM-Pro grant 05A2023.  LP acknowledges support from NSFC under grant No.\ 11973098 and No.\ 12373072. The authors also gratefully acknowledge the Gauss Centre for Supercomputing e.V. (www.gauss-centre.eu) for providing computing time through the John von Neumann Institute for Computing (NIC) on the GCS Supercomputer JUWELS at J\"ulich Supercomputing Centre (JSC). We thank the anonymous referee for their careful reading of our manuscript and for their constructive suggestions.  ChatGPT (version GPT-5; OpenAI) was used in formatting the tables in this manuscript.

\bibliographystyle{aasjournal}
\bibliography{Turb.bib}

\section{Appendix}

In Tables  \ref{tab:momentsM_C} and  \ref{tab:momentsM_MS}, we give the full data for the moments of $P_{\rm M}(s),$ as discussed in \S \ref{sec:mass}.  Here, the derived quantities have been normalized by the mass-weighted Mach number, $M_{\rm M}$ matching the normalization in Fig.\ \ref{fig:fitMM}.

\begin{figure*}[t]
\begin{center}
\includegraphics[width=\textwidth]{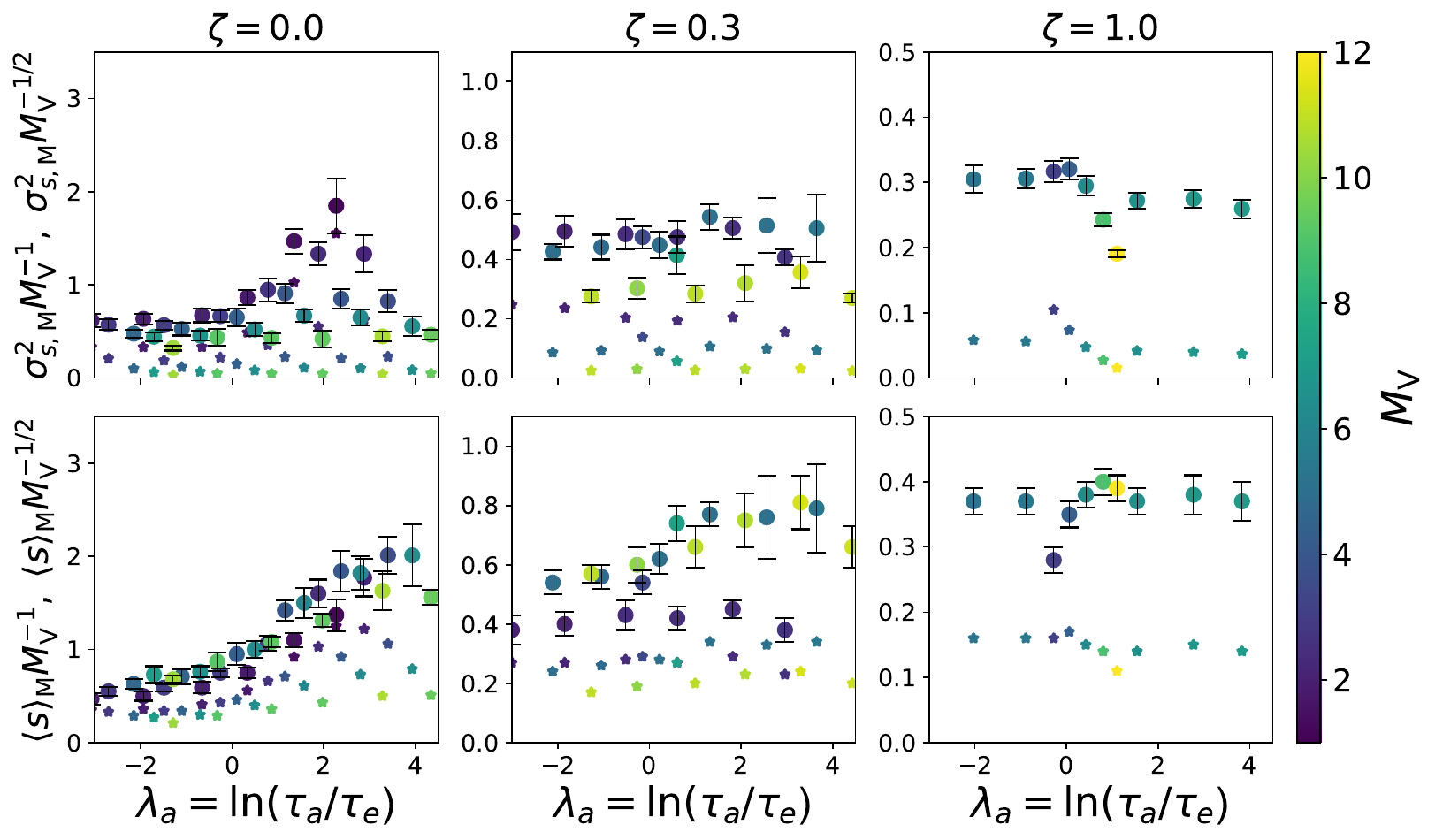}
\end{center}
\vspace{-0.1in}
\caption{{\em Top:} The mass-weighted variance of $s$, normalized by the volume-weighted Mach number, $ \sigma^2_{s,{\rm M}} M_{\rm V}^{-1},$ (stars) and by the square root of the volume-weighted Mach number, $\sigma^2_{s,{\rm M}} M_{\rm V}^{-1/2},$ (circles). Columns show results from compressively-driven turbulence (left), mixed-driving turbulence (center), and solenoidally-driven turbulence (right).  {\em Bottom:} The mass-weighted variance mean values of $s$, normalized by the volume-weighted Mach number, $\left< s \right>_{\rm M} M_{\rm V}^{-1},$ (stars) and by the square root of the volume-weighted Mach number, $\left< s \right>_{\rm M} M_{\rm V}^{-1/2},$ (stars).}
\label{fig:fitVM}
\end{figure*}

Fig.\ \ref{fig:fitVM} shows the results of normalizing the derived mass-weighted quantities by the volume-weighted rms Mach number. While the details change, the trends do not. Furthermore, the scatter in the various quantities is similar to that seen in Fig.\ \ref{fig:fitMM}.  As there is no strong statistical reason to prefer one normalization to another, we choose to work with $M_{\ rm M} $ in our study, as in this case, the normalized mass-weighted moments are constructed from purely mass-weighted quantities.

\begin{table*}[t]
\hspace{-0.5in}
 \centering
 \resizebox{1.0\textwidth}{!}{
\begin{tabular}{lccccccc}
\toprule
 & \multicolumn{3}{c}{PDF Moments} & \multicolumn{4}{c}{Derived Values}\\
\cmidrule(lr{.75em}){2-4} \cmidrule(lr{.75em}){5-8}
Name & $\left<s\right>_{\rm M}$ & $\sigma_{\rm M}^2$ & $\mu_{\rm M}$ & $\sigma_{\rm M}^2 M_{\rm M}^{-1}$ & $\sigma_{\rm M}^2 M_{\rm M}^{-1/2}$ & $\left<s\right>_{\rm M} M_{\rm M}^{-1}$ & $\left<s\right>_{\rm M} M_{\rm M}^{-1/2}$ \\
$\mathtt{Ms1.9\_C\_\lambda-3.1}$ & 0.64 $\pm$ 0.07 & 1.15 $\pm$ 0.13 & -0.28 $\pm$ 0.15 & 0.63 $\pm$ 0.09 & 0.85 $\pm$ 0.10 & 0.35 $\pm$ 0.05 & 0.48 $\pm$ 0.06 \\
$\mathtt{Ms1.9\_C\_\lambda-1.9}$ & 0.69 $\pm$ 0.06 & 1.22 $\pm$ 0.11 & -0.31 $\pm$ 0.18 & 0.67 $\pm$ 0.09 & 0.90 $\pm$ 0.09 & 0.38 $\pm$ 0.05 & 0.51 $\pm$ 0.05 \\
$\mathtt{Ms2.0\_C\_\lambda-0.7}$ & 0.84 $\pm$ 0.07 & 1.36 $\pm$ 0.15 & -0.45 $\pm$ 0.13 & 0.77 $\pm$ 0.11 & 1.03 $\pm$ 0.12 & 0.47 $\pm$ 0.06 & 0.63 $\pm$ 0.06 \\
$\mathtt{Ms1.8\_C\_\lambda+0.3}$ & 0.99 $\pm$ 0.08 & 1.53 $\pm$ 0.14 & -0.57 $\pm$ 0.12 & 1.09 $\pm$ 0.17 & 1.29 $\pm$ 0.13 & 0.71 $\pm$ 0.11 & 0.84 $\pm$ 0.07 \\
$\mathtt{Ms1.4\_C\_\lambda+1.4}$ & 1.31 $\pm$ 0.09 & 2.10 $\pm$ 0.19 & -0.50 $\pm$ 0.13 & 2.19 $\pm$ 0.34 & 2.14 $\pm$ 0.21 & 1.37 $\pm$ 0.20 & 1.34 $\pm$ 0.10 \\
$\mathtt{Ms1.2\_C\_\lambda+2.3}$ & 1.49 $\pm$ 0.16 & 2.20 $\pm$ 0.35 & -0.65 $\pm$ 0.15 & 2.97 $\pm$ 1.14 & 2.56 $\pm$ 0.43 & 2.01 $\pm$ 0.73 & 1.73 $\pm$ 0.22 \\
[0.3em]
$\mathtt{Ms2.8\_C\_\lambda-2.7}$ & 0.92 $\pm$ 0.07 & 1.58 $\pm$ 0.16 & -0.31 $\pm$ 0.14 & 0.60 $\pm$ 0.09 & 0.97 $\pm$ 0.11 & 0.35 $\pm$ 0.05 & 0.56 $\pm$ 0.06 \\
$\mathtt{Ms3.0\_C\_\lambda-1.5}$ & 1.03 $\pm$ 0.06 & 1.70 $\pm$ 0.14 & -0.36 $\pm$ 0.13 & 0.62 $\pm$ 0.08 & 1.02 $\pm$ 0.10 & 0.37 $\pm$ 0.04 & 0.62 $\pm$ 0.05 \\
$\mathtt{Ms3.0\_C\_\lambda-0.3}$ & 1.31 $\pm$ 0.08 & 2.00 $\pm$ 0.19 & -0.47 $\pm$ 0.13 & 0.80 $\pm$ 0.11 & 1.26 $\pm$ 0.13 & 0.52 $\pm$ 0.06 & 0.83 $\pm$ 0.06 \\
$\mathtt{Ms2.7\_C\_\lambda+0.8}$ & 1.76 $\pm$ 0.11 & 2.54 $\pm$ 0.33 & -0.56 $\pm$ 0.20 & 1.32 $\pm$ 0.28 & 1.83 $\pm$ 0.27 & 0.91 $\pm$ 0.17 & 1.27 $\pm$ 0.10 \\
$\mathtt{Ms2.4\_C\_\lambda+1.9}$ & 2.49 $\pm$ 0.07 & 3.23 $\pm$ 0.29 & -0.81 $\pm$ 0.14 & 2.53 $\pm$ 1.58 & 2.86 $\pm$ 0.28 & 1.95 $\pm$ 1.21 & 2.20 $\pm$ 0.20 \\
$\mathtt{Ms2.1\_C\_\lambda+2.9}$ & 2.59 $\pm$ 0.20 & 2.84 $\pm$ 0.43 & -0.96 $\pm$ 0.17 & 2.63 $\pm$ 1.68 & 2.73 $\pm$ 0.45 & 2.40 $\pm$ 1.50 & 2.49 $\pm$ 0.28 \\
[0.3em]
$\mathtt{Ms4.7\_C\_\lambda-2.1}$ & 1.37 $\pm$ 0.08 & 2.23 $\pm$ 0.19 & -0.37 $\pm$ 0.10 & 0.51 $\pm$ 0.06 & 1.06 $\pm$ 0.11 & 0.31 $\pm$ 0.03 & 0.65 $\pm$ 0.05 \\
$\mathtt{Ms4.4\_C\_\lambda-1.1}$ & 1.50 $\pm$ 0.15 & 2.34 $\pm$ 0.31 & -0.43 $\pm$ 0.18 & 0.60 $\pm$ 0.10 & 1.18 $\pm$ 0.16 & 0.38 $\pm$ 0.05 & 0.76 $\pm$ 0.08 \\
$\mathtt{Ms4.3\_C\_\lambda+0.1}$ & 1.98 $\pm$ 0.23 & 2.82 $\pm$ 0.40 & -0.56 $\pm$ 0.15 & 0.82 $\pm$ 0.19 & 1.52 $\pm$ 0.23 & 0.57 $\pm$ 0.12 & 1.07 $\pm$ 0.14 \\
$\mathtt{Ms4.0\_C\_\lambda+1.2}$ & 2.85 $\pm$ 0.10 & 3.64 $\pm$ 0.41 & -0.72 $\pm$ 0.21 & 1.47 $\pm$ 0.54 & 2.31 $\pm$ 0.28 & 1.15 $\pm$ 0.41 & 1.81 $\pm$ 0.14 \\
$\mathtt{Ms4.0\_C\_\lambda+2.4}$ & 3.69 $\pm$ 0.29 & 3.41 $\pm$ 0.41 & -1.17 $\pm$ 0.18 & 2.02 $\pm$ 2.00 & 2.62 $\pm$ 0.33 & 2.19 $\pm$ 2.16 & 2.84 $\pm$ 0.34 \\
$\mathtt{Ms3.6\_C\_\lambda+3.4}$ & 3.81 $\pm$ 0.20 & 2.97 $\pm$ 0.42 & -1.18 $\pm$ 0.17 & 2.38 $\pm$ 3.30 & 2.66 $\pm$ 0.39 & 3.05 $\pm$ 4.22 & 3.41 $\pm$ 0.33 \\
[0.3em]
$\mathtt{Ms7.1\_C\_\lambda-1.7}$ & 1.94 $\pm$ 0.19 & 3.14 $\pm$ 0.34 & -0.37 $\pm$ 0.13 & 0.45 $\pm$ 0.08 & 1.19 $\pm$ 0.17 & 0.28 $\pm$ 0.05 & 0.74 $\pm$ 0.09 \\
$\mathtt{Ms6.6\_C\_\lambda-0.7}$ & 1.95 $\pm$ 0.13 & 3.01 $\pm$ 0.34 & -0.36 $\pm$ 0.13 & 0.53 $\pm$ 0.09 & 1.26 $\pm$ 0.16 & 0.34 $\pm$ 0.05 & 0.82 $\pm$ 0.06 \\
$\mathtt{Ms6.3\_C\_\lambda+0.5}$ & 2.53 $\pm$ 0.19 & 3.31 $\pm$ 0.47 & -0.55 $\pm$ 0.19 & 0.69 $\pm$ 0.16 & 1.51 $\pm$ 0.23 & 0.53 $\pm$ 0.10 & 1.15 $\pm$ 0.10 \\
$\mathtt{Ms6.1\_C\_\lambda+1.6}$ & 3.70 $\pm$ 0.19 & 4.06 $\pm$ 0.39 & -0.84 $\pm$ 0.14 & 1.09 $\pm$ 0.55 & 2.10 $\pm$ 0.22 & 0.99 $\pm$ 0.50 & 1.91 $\pm$ 0.21 \\
$\mathtt{Ms6.2\_C\_\lambda+2.8}$ & 4.54 $\pm$ 0.25 & 4.05 $\pm$ 0.54 & -1.11 $\pm$ 0.19 & 1.44 $\pm$ 1.16 & 2.41 $\pm$ 0.36 & 1.61 $\pm$ 1.29 & 2.71 $\pm$ 0.27 \\
$\mathtt{Ms6.5\_C\_\lambda+3.9}$ & 5.11 $\pm$ 0.26 & 3.60 $\pm$ 0.70 & -1.32 $\pm$ 0.18 & 1.96 $\pm$ 7.58 & 2.66 $\pm$ 0.53 & 2.78 $\pm$ 10.75 & 3.77 $\pm$ 0.62 \\
[0.3em]
$\mathtt{Ms10.4\_C\_\lambda-1.3}$ & 2.20 $\pm$ 0.10 & 3.36 $\pm$ 0.29 & -0.42 $\pm$ 0.11 & 0.34 $\pm$ 0.04 & 1.07 $\pm$ 0.11 & 0.22 $\pm$ 0.02 & 0.70 $\pm$ 0.04 \\
$\mathtt{Ms9.1\_C\_\lambda-0.3}$ & 2.62 $\pm$ 0.29 & 3.99 $\pm$ 0.79 & -0.40 $\pm$ 0.14 & 0.52 $\pm$ 0.11 & 1.44 $\pm$ 0.29 & 0.34 $\pm$ 0.05 & 0.95 $\pm$ 0.11 \\
$\mathtt{Ms9.2\_C\_\lambda+0.9}$ & 3.29 $\pm$ 0.09 & 3.98 $\pm$ 0.47 & -0.56 $\pm$ 0.18 & 0.59 $\pm$ 0.12 & 1.53 $\pm$ 0.19 & 0.48 $\pm$ 0.08 & 1.26 $\pm$ 0.07 \\
$\mathtt{Ms9.3\_C\_\lambda+2.0}$ & 4.01 $\pm$ 0.19 & 3.93 $\pm$ 0.84 & -0.83 $\pm$ 0.29 & 0.67 $\pm$ 0.17 & 1.63 $\pm$ 0.37 & 0.69 $\pm$ 0.09 & 1.66 $\pm$ 0.09 \\
$\mathtt{Ms10.6\_C\_\lambda+3.3}$ & 5.30 $\pm$ 0.31 & 4.74 $\pm$ 0.56 & -1.12 $\pm$ 0.21 & 1.17 $\pm$ 1.78 & 2.35 $\pm$ 0.30 & 1.31 $\pm$ 1.98 & 2.63 $\pm$ 0.33 \\
$\mathtt{Ms9.4\_C\_\lambda+4.3}$ & 4.78 $\pm$ 0.11 & 4.39 $\pm$ 0.46 & -0.97 $\pm$ 0.16 & 0.98 $\pm$ 0.40 & 2.07 $\pm$ 0.22 & 1.06 $\pm$ 0.43 & 2.25 $\pm$ 0.11 \\
\bottomrule
\end{tabular}}
\vspace{0.2in}
\caption{Mass-weighted moments and derived quantities from our compressively-driven simulations. }
\label{tab:momentsM_C}
\end{table*}

\begin{table*}[t]
\hspace{-0.5in}
 \centering
 \resizebox{1.0\textwidth}{!}{
\begin{tabular}{lccccccc}
\toprule
 & \multicolumn{3}{c}{PDF Moments} & \multicolumn{4}{c}{Derived Values}\\
\cmidrule(lr{.75em}){2-4} \cmidrule(lr{.75em}){5-8}
Name & $\left<s\right>_{\rm M}$ & $\sigma_{\rm M}^2$ & $\mu_{\rm M}$ & $\sigma_{\rm M}^2 M_{\rm M}^{-1}$ & $\sigma_{\rm M}^2 M_{\rm M}^{-1/2}$ & $\left<s\right>_{\rm M} M_{\rm M}^{-1}$ & $\left<s\right>_{\rm M} M_{\rm M}^{-1/2}$ \\
$\mathtt{Ms2.0\_M\_\lambda-3.0}$ & 0.54 $\pm$ 0.06 & 0.98 $\pm$ 0.12 & -0.24 $\pm$ 0.13 & 0.51 $\pm$ 0.07 & 0.71 $\pm$ 0.09 & 0.28 $\pm$ 0.04 & 0.39 $\pm$ 0.05 \\
$\mathtt{Ms2.1\_M\_\lambda-1.8}$ & 0.58 $\pm$ 0.06 & 1.04 $\pm$ 0.11 & -0.28 $\pm$ 0.13 & 0.52 $\pm$ 0.07 & 0.74 $\pm$ 0.08 & 0.29 $\pm$ 0.04 & 0.41 $\pm$ 0.04 \\
$\mathtt{Ms2.4\_M\_\lambda-0.5}$ & 0.67 $\pm$ 0.07 & 1.16 $\pm$ 0.12 & -0.33 $\pm$ 0.14 & 0.54 $\pm$ 0.07 & 0.79 $\pm$ 0.09 & 0.31 $\pm$ 0.04 & 0.45 $\pm$ 0.05 \\
$\mathtt{Ms2.5\_M\_\lambda+0.6}$ & 0.67 $\pm$ 0.05 & 1.17 $\pm$ 0.13 & -0.35 $\pm$ 0.15 & 0.56 $\pm$ 0.08 & 0.81 $\pm$ 0.09 & 0.32 $\pm$ 0.04 & 0.46 $\pm$ 0.04 \\
$\mathtt{Ms2.5\_M\_\lambda+1.8}$ & 0.71 $\pm$ 0.04 & 1.25 $\pm$ 0.09 & -0.33 $\pm$ 0.13 & 0.63 $\pm$ 0.08 & 0.89 $\pm$ 0.07 & 0.36 $\pm$ 0.04 & 0.50 $\pm$ 0.03 \\
$\mathtt{Ms2.6\_M\_\lambda+3.0}$ & 0.61 $\pm$ 0.05 & 1.07 $\pm$ 0.07 & -0.37 $\pm$ 0.10 & 0.46 $\pm$ 0.08 & 0.70 $\pm$ 0.07 & 0.27 $\pm$ 0.04 & 0.40 $\pm$ 0.04 \\
$\mathtt{Ms3.5\_M\_\lambda-0.2}$ & 1.00 $\pm$ 0.07 & 1.64 $\pm$ 0.13 & -0.40 $\pm$ 0.12 & 0.55 $\pm$ 0.06 & 0.95 $\pm$ 0.08 & 0.34 $\pm$ 0.04 & 0.58 $\pm$ 0.04 \\
$\mathtt{Ms5.0\_M\_\lambda-2.1}$ & 1.21 $\pm$ 0.08 & 2.11 $\pm$ 0.13 & -0.26 $\pm$ 0.10 & 0.45 $\pm$ 0.04 & 0.97 $\pm$ 0.07 & 0.26 $\pm$ 0.02 & 0.56 $\pm$ 0.04 \\
$\mathtt{Ms4.8\_M\_\lambda-1.0}$ & 1.23 $\pm$ 0.08 & 2.12 $\pm$ 0.20 & -0.25 $\pm$ 0.13 & 0.49 $\pm$ 0.07 & 1.02 $\pm$ 0.10 & 0.28 $\pm$ 0.03 & 0.59 $\pm$ 0.04 \\
$\mathtt{Ms5.0\_M\_\lambda+0.2}$ & 1.38 $\pm$ 0.08 & 2.24 $\pm$ 0.22 & -0.35 $\pm$ 0.15 & 0.55 $\pm$ 0.09 & 1.11 $\pm$ 0.11 & 0.33 $\pm$ 0.05 & 0.68 $\pm$ 0.05 \\
$\mathtt{Ms5.1\_M\_\lambda+1.3}$ & 1.74 $\pm$ 0.07 & 2.79 $\pm$ 0.22 & -0.44 $\pm$ 0.10 & 0.75 $\pm$ 0.13 & 1.44 $\pm$ 0.12 & 0.47 $\pm$ 0.07 & 0.90 $\pm$ 0.05 \\
$\mathtt{Ms5.2\_M\_\lambda+2.6}$ & 1.74 $\pm$ 0.31 & 2.67 $\pm$ 0.48 & -0.50 $\pm$ 0.13 & 0.67 $\pm$ 0.15 & 1.34 $\pm$ 0.27 & 0.44 $\pm$ 0.10 & 0.87 $\pm$ 0.16 \\
$\mathtt{Ms5.4\_M\_\lambda+3.7}$ & 1.83 $\pm$ 0.33 & 2.73 $\pm$ 0.61 & -0.54 $\pm$ 0.19 & 0.68 $\pm$ 0.24 & 1.36 $\pm$ 0.34 & 0.46 $\pm$ 0.15 & 0.92 $\pm$ 0.18 \\
$\mathtt{Ms7.3\_M\_\lambda+0.6}$ & 1.99 $\pm$ 0.13 & 3.03 $\pm$ 0.46 & -0.46 $\pm$ 0.17 & 0.52 $\pm$ 0.12 & 1.26 $\pm$ 0.21 & 0.34 $\pm$ 0.06 & 0.83 $\pm$ 0.07 \\
$\mathtt{Ms10.9\_M\_\lambda-1.3}$ & 1.89 $\pm$ 0.09 & 3.00 $\pm$ 0.23 & -0.38 $\pm$ 0.09 & 0.29 $\pm$ 0.03 & 0.93 $\pm$ 0.08 & 0.18 $\pm$ 0.01 & 0.58 $\pm$ 0.03 \\
$\mathtt{Ms10.1\_M\_\lambda-0.3}$ & 1.92 $\pm$ 0.19 & 3.07 $\pm$ 0.37 & -0.36 $\pm$ 0.08 & 0.32 $\pm$ 0.05 & 0.99 $\pm$ 0.14 & 0.20 $\pm$ 0.03 & 0.62 $\pm$ 0.07 \\
$\mathtt{Ms10.9\_M\_\lambda+1.0}$ & 2.16 $\pm$ 0.21 & 3.09 $\pm$ 0.31 & -0.47 $\pm$ 0.12 & 0.36 $\pm$ 0.06 & 1.06 $\pm$ 0.12 & 0.25 $\pm$ 0.04 & 0.74 $\pm$ 0.08 \\
$\mathtt{Ms10.7\_M\_\lambda+2.1}$ & 2.45 $\pm$ 0.30 & 3.44 $\pm$ 0.66 & -0.53 $\pm$ 0.13 & 0.43 $\pm$ 0.09 & 1.22 $\pm$ 0.24 & 0.31 $\pm$ 0.05 & 0.87 $\pm$ 0.11 \\
$\mathtt{Ms11.4\_M\_\lambda+3.3}$ & 2.74 $\pm$ 0.24 & 4.05 $\pm$ 0.60 & -0.45 $\pm$ 0.16 & 0.53 $\pm$ 0.19 & 1.47 $\pm$ 0.23 & 0.36 $\pm$ 0.12 & 0.99 $\pm$ 0.11 \\
$\mathtt{Ms11.2\_M\_\lambda+4.4}$ & 2.20 $\pm$ 0.13 & 3.01 $\pm$ 0.17 & -0.50 $\pm$ 0.09 & 0.37 $\pm$ 0.11 & 1.05 $\pm$ 0.08 & 0.27 $\pm$ 0.08 & 0.77 $\pm$ 0.08 \\[0.3em]
$\mathtt{Ms3.0\_S\_\lambda-0.3}$ & 0.49 $\pm$ 0.02 & 0.96 $\pm$ 0.05 & -0.08 $\pm$ 0.07 & 0.34 $\pm$ 0.03 & 0.57 $\pm$ 0.03 & 0.17 $\pm$ 0.01 & 0.29 $\pm$ 0.02 \\$\mathtt{Ms4.4\_S\_\lambda+0.1}$ & 0.73 $\pm$ 0.04 & 1.40 $\pm$ 0.07 & -0.14 $\pm$ 0.06 & 0.34 $\pm$ 0.02 & 0.69 $\pm$ 0.04 & 0.18 $\pm$ 0.01 & 0.36 $\pm$ 0.02 \\
$\mathtt{Ms5.3\_S\_\lambda-2.0}$ & 0.84 $\pm$ 0.04 & 1.60 $\pm$ 0.11 & -0.15 $\pm$ 0.06 & 0.32 $\pm$ 0.03 & 0.72 $\pm$ 0.05 & 0.17 $\pm$ 0.02 & 0.38 $\pm$ 0.02 \\
$\mathtt{Ms5.5\_S\_\lambda-0.9}$ & 0.87 $\pm$ 0.04 & 1.67 $\pm$ 0.08 & -0.12 $\pm$ 0.06 & 0.32 $\pm$ 0.02 & 0.73 $\pm$ 0.04 & 0.17 $\pm$ 0.01 & 0.38 $\pm$ 0.02 \\
$\mathtt{Ms6.2\_S\_\lambda+0.4}$ & 0.96 $\pm$ 0.04 & 1.84 $\pm$ 0.09 & -0.14 $\pm$ 0.05 & 0.31 $\pm$ 0.02 & 0.76 $\pm$ 0.04 & 0.16 $\pm$ 0.01 & 0.40 $\pm$ 0.02 \\
$\mathtt{Ms6.5\_S\_\lambda+1.5}$ & 0.94 $\pm$ 0.05 & 1.78 $\pm$ 0.08 & -0.15 $\pm$ 0.06 & 0.29 $\pm$ 0.02 & 0.72 $\pm$ 0.04 & 0.15 $\pm$ 0.01 & 0.38 $\pm$ 0.02 \\
$\mathtt{Ms6.9\_S\_\lambda+2.8}$ & 1.00 $\pm$ 0.05 & 1.89 $\pm$ 0.09 & -0.15 $\pm$ 0.07 & 0.29 $\pm$ 0.03 & 0.74 $\pm$ 0.04 & 0.16 $\pm$ 0.02 & 0.40 $\pm$ 0.03 \\
$\mathtt{Ms7.0\_S\_\lambda+3.9}$ & 0.97 $\pm$ 0.04 & 1.83 $\pm$ 0.10 & -0.14 $\pm$ 0.07 & 0.27 $\pm$ 0.04 & 0.70 $\pm$ 0.05 & 0.14 $\pm$ 0.02 & 0.37 $\pm$ 0.03 \\
$\mathtt{Ms8.9\_S\_\lambda+0.8}$  & 1.20 $\pm$ 0.06 & 2.15 $\pm$ 0.09 & -0.23 $\pm$ 0.05 & 0.25 $\pm$ 0.02 & 0.74 $\pm$ 0.04 & 0.14 $\pm$ 0.01 & 0.41 $\pm$ 0.03 \\
$\mathtt{Ms12.3\_S\_\lambda+1.2}$ & 1.35 $\pm$ 0.06 & 2.34 $\pm$ 0.07 & -0.29 $\pm$ 0.05 & 0.19 $\pm$ 0.02 & 0.67 $\pm$ 0.04 & 0.11 $\pm$ 0.01 & 0.39 $\pm$ 0.02 \\
\bottomrule
\end{tabular}}
\vspace{0.2in}
\caption{Mass-weighted moments and derived quantities from our mixed and solenoidally-driven simulations. }
\label{tab:momentsM_MS}
\end{table*}%

\end{document}